\documentclass{emulateapj}
\usepackage{apjfonts}
\usepackage{mathptmx}

\newcommand{\eg}{e.g.,}

\newcommand{\etal}{et~al.\ }
\newcommand{\ltsima}{$\; \buildrel < \over \sim \;$}
\newcommand{\simlt}{\lower.5ex\hbox{\ltsima}}
\newcommand{\gtsima}{$\; \buildrel > \over \sim \;$}
\newcommand{\simgt}{\lower.5ex\hbox{\gtsima}}
\newcommand{\kms}{km s$^{-1}$}
\newcommand{\ho}{km~s$^{-1}$~Mpc$^{-1}$}
 
\newcommand{\dm}{$\Delta m_{15}(B)$}

\def\parcmin{{\tt '}\mskip -6.0mu.\,}
\def\parcsec{{\tt ''}\mskip -6.0mu.\,}

\begin{document}
 
\title{Calibrating Type Ia Supernovae using the 
Planetary Nebula Luminosity Function I. Initial Results}

\author{John J. Feldmeier\altaffilmark{1,2,3}}  
\affil{Department of Physics \& Astronomy, Youngstown State
University, Youngstown, OH 44555-2001}
\email{jjfeldmeier@ysu.edu}

\author{George H. Jacoby}
\affil{WIYN Observatory
\\ P.O. Box 26732, Tucson, AZ 85726}
\email{gjacoby@wiyn.org}

\author{M. M. Phillips}
\affil{Las Campanas Observatory, Carnegie Observatories, Casilla 601, 
La Serena, Chile}
\email{mmp@lco.cl}

\altaffiltext{1}{The WIYN Observatory is a joint facility of the University 
of Wisconsin-Madison, Indiana University, Yale University, and the 
National Optical Astronomy Observatory.}

\altaffiltext{2}{This paper includes data gathered with the 6.5 meter 
Magellan Telescopes located at Las Campanas Observatory, Chile}

\altaffiltext{3}{NSF Astronomy and Astrophysics Postdoctoral Fellow}

\begin{abstract}

We report the results of an [O~III] $\lambda 5007$ survey for planetary
nebulae (PN) in five galaxies that were hosts of well-observed Type~Ia 
supernovae: NGC~524, NGC~1316, NGC~1380, NGC~1448 and NGC~4526.  
The goals of this
survey are to better quantify the zero-point of the maximum magnitude
versus decline rate relation for supernovae 
Type~Ia and to validate the insensitivity
of Type~Ia luminosity to parent stellar population using the 
host galaxy Hubble type as a surrogate.  We detected a total of 45 
planetary nebulae candidates in NGC~1316, 44 candidates in NGC~1380, and
94 candidates in NGC~4526.  From these data, and the empirical planetary 
nebula luminosity function (PNLF), we derive distances of 
$17.9^{+0.8}_{-0.9}$~Mpc, $16.1^{+0.8}_{-1.1}$~Mpc, and
$13.6^{+1.3}_{-1.2}$~Mpc respectively.  Our derived distance to NGC~4526
has a lower precision due to the likely presence of
Virgo intracluster planetary nebulae in the foreground of this galaxy.  
In NGC~524 and NGC~1448 we detected no planetary 
nebulae candidates down to the limiting magnitudes of our observations.  
We present a formalism for setting realistic distance limits in these 
two cases, and derive robust lower limits of 20.9~Mpc and 15.8~Mpc, 
respectively.  

After combining these results with other distances from
the PNLF, Cepheid, and Surface Brightness Fluctuations distance indicators,
we calibrate the optical and near-infrared relations for supernovae Type~Ia
and we find that the Hubble constants derived from each of the three 
methods are broadly consistent, implying that the properties of 
supernovae Type~Ia do not
vary drastically as a function of stellar population.  We determine a 
preliminary Hubble constant of H$_{0}$ =  77 $\pm$ 3 (random) $\pm$ 5 (systematic)
\ho for the PNLF, though more nearby galaxies with high-quality 
observations are clearly needed.
\end{abstract}
 
 
\section{Introduction}

Type~Ia supernovae (SNe Ia) have become one of the 
best ways for determining the Hubble constant, the deceleration 
parameter, and the cosmological constant (for a recent review, see 
Filippenko 2005).  They are the most luminous of all supernovae, 
and can be observed to high redshifts ($z \approx 1.6$; 
see Strolger \& Riess (2006), 
and references therein).  Although SNe~Ia are known to vary in
maximum brightness by up to a magnitude in the $B$ band 
%at maximum light 
\citep{hamuy1996a}, there is a tight correlation between the decline
rate of the light curve and the maximum luminosity 
\citep{phillips1993,hamuy1996a,phillips1999}.  With corrections for this
effect \citep{hamuy1996b,riess1996,perlmutter1999}, the relative 
dispersion in SNe~Ia maximum magnitudes, and therefore the relative
distances, is less than 0.2 magnitudes.  This high precision has allowed
the detection of the acceleration of the universe 
\citep{riess1998,perlmutter1999}.

However, it should be stressed that in spite of the impressive 
consistency seen in the relative magnitudes of SNe~Ia, universal 
agreement on the absolute magnitudes has not been achieved.  
The SNe~Ia zero-point calibrations of \citet{gibson2000}, 
\citet{saha2001,saha2006} and \citet{freedman2001} lead to values 
that vary by $\approx 0.3$ magnitudes.  
Remarkably, these determinations are based on the identical {\sl HST} 
observations of Cepheids in galaxies that hosted SNe~Ia, 
with the same assumed distance modulus to the Large Magellanic Cloud.
The differences can be attributed to the use of different Cepheid samples
and Period-Luminosity relationships, and different assumptions concerning
metallicity corrections and which historical Type~Ia supernovae to include
or exclude in these relations.   

There are further complications with relying solely on the current
{\sl HST} Cepheid zero-point calibration of SNe~Ia's.
First, approximately 27\% of SNe~Ia are hosted in early-type
galaxies \citep{barbon1999}.  By limiting the zero-point determination
to only galaxies that host with Cepheid stars, we are throwing
away approximately a quarter of nearby supernovae, a dubious 
luxury at best, especially considering how few Type~Ia supernovae
are within range of direct distance indicators.  

Second, and more importantly, observing the absolute 
magnitudes of SNe~Ia only over a small range of stellar populations 
limits our ability to test for any evolutionary differences that
might appear in high redshift samples.  Theoretical studies give mixed 
results on the amount and sign of any evolutionary effect on the 
maximum magnitude of SNe~Ia \citep{hoeflich1998,umeda1999,nomoto1999,yungelson2000}, 
but the effect could be significant, up to an additional 0.3 magnitudes. 

These theoretical analyses are supported by observational evidence
showing that there are clear differences between the observed magnitudes
of SNe~Ia observed in elliptical galaxies and SNe~Ia observed in spiral galaxies.  
Early results by Hamuy \etal (1995) and follow-up studies 
\citep{hamuy1996b,hamuy2000,gallagher2005} clearly show that SNe~Ia 
in early-type galaxies are preferentially less luminous, and have a faster 
decline rate than those found in late-type galaxies.  More recently, 
\citet{sullivan2006} has shown that these trends continue to redshifts 
as large as 0.75.  \citet{gallagher2005} has given evidence implying 
that the stellar population age is the driving force behind the different 
properties of SNe~Ia.   

It is an open question whether these differences between the supernovae hosted by 
early type and late type galaxies are fully corrected by the decline rate correction, 
or whether there might be additional systematic effects left to be uncovered. There
is now some tentative evidence for the latter view.  \citet{gallagher2005} 
has reported a possible negative correlation between host galaxy metallicity and 
Hubble residual, after correction for decline rate.  Although the 
measured difference is less than 2$\sigma$ in magnitude, this result, if 
correct, would imply that metallicity could change the maximum brightness 
of SNe~Ia up to 10\%.

Consequently, it may be beneficial to derive distances to galaxies
that have hosted SNe~Ia and contain more diverse stellar 
populations than the Cepheid distance indicator can probe.
In particular, the [O~III] $\lambda$ 5007 
Planetary Nebulae Luminosity Function (PNLF) distance indicator
\citep{mudville,rbc2005} is well-suited to measure the zero-point of SNe~Ia.
The scatter of the PNLF method is comparable to the scatter of Cepheid 
distance determinations \citep[][]{pnlf12}, making the PNLF 
an equally precise distance
indicator.  Since planetary nebulae are present in galaxies of all 
Hubble types, the PNLF method can be used in all galaxies that host 
SNe~Ia, allowing for tests of evolutionary effects.

Given these advantages, we have begun a program to measure distances 
to early and late type galaxies that were hosts to well-observed 
SNe~Ia using the PNLF distance indicator.  With the larger 
aperture of the 6.5m Magellan Clay telescope and the excellent
seeing of the 3.5m WIYN telescope, we can extend the reach of 
the PNLF method significantly further than the $\approx$ 17 Mpc 
limit previously reached with 
4-m class telescopes.  Our primary goal is to increase the sample of
SNe~Ia with high quality distances within 20 Mpc.  In this
paper, we present the first results of our program, and give
a revised Hubble constant from the PNLF results thus far.    

\section{Observations and Reductions}

For this first study, we chose five galaxies that have hosted 
well-observed SNe~Ia
in the past twenty five years: NGC~524, NGC~1316, NGC~1380, 
NGC~1448, and NGC~4526.  Most of these galaxies are early-type 
spirals and lenticulars,
being unsuitable for Cepheid observations.  Properties of the target 
galaxies, and the observed supernovae that they hosted are given 
in Table~\ref{table:target}.  

Two telescopes were used in this study.  Our first 
observations were obtained on 18--20 December 2003 and 11--12 November
2004 using the Magellan Landon Clay telescope, and the MagIC camera 
\citep{osip2004}.  This camera consists of a SITe 2048x2048 CCD detector 
with 24 $\mu$m pixels,
and uses four amplifiers, with a mean gain of 1.96 electrons per ADU, 
and a mean read noise of 5.25 electrons.  This instrumental setup gave 
a pixel scale of 0.069 arcseconds per pixel and 
a field-of-view of $2\parcmin36 \times 2\parcmin36$.  We supplemented
our Magellan observations with additional data obtained on 10--11 March
2005 using the WIYN 3.5m telescope, and the OPTIC camera 
\citep{howell2003,tonry2002}.  This camera consists of two 2Kx4K  
CCID-28 orthogonal transfer CCDs arranged in a single dewar.  This
setup has a read noise of 4 electrons and a gain of 1.45 electrons per ADU. 
The instrumental set-up at WIYN gave a pixel scale of 0.14 arcseconds
per pixel and a total field-of-view of $9\parcmin56 \times 9\parcmin56$.

\begin{figure}
\figurenum{1}
\label{fig:filters}
\plotone{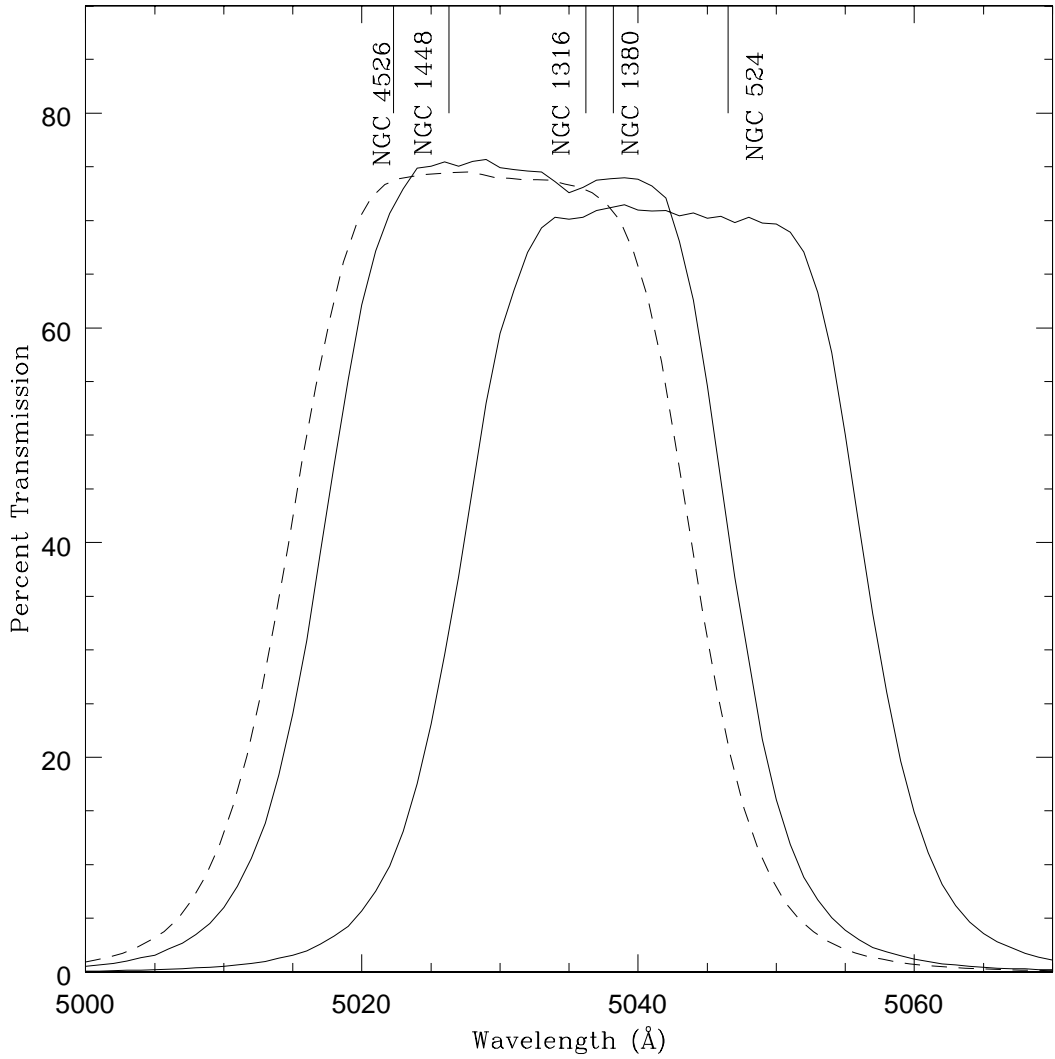}
\caption{A plot of the transmission profiles of the two [O~III] 
$\lambda$ 5007 filters used in our observations.  The solid lines
indicate the bandpasses at the Magellan telescope, while the dashed
line indicates the bandpass at the WIYN telescope.  The mean-redshifted
position of the 5007 line is shown for each galaxy as the vertical 
lines at the top of the figure.  In the case of NGC~4526, we focused
on the red-shifted side of the galaxy, and that location is given here.}  
\end{figure}

Our survey technique is similar to other extragalactic searches for 
planetary nebulae \citep{pnlf12}.  Depending on redshift, 
we obtained exposures for our target galaxies through one of two 
30~\AA\ wide [O~III] $\lambda 5007$ filters, whose central wavelengths
are 5027~\AA\ and 5040~\AA\, respectively.  The filter curves, compared
against the expected wavelength of the red-shifted [O~III] $\lambda$ 5007
emission line are shown in Figure~\ref{fig:filters}.  Corresponding 
images were then 
taken for all the galaxies through a 230~\AA\ wide off-band 
continuum filter 
(central wavelength $\sim 5288$~\AA).  In the case of NGC~1448 and
NGC~4526,
exposures were also taken through a 44~\AA\ wide H$\alpha$ filter
(central wavelength $\sim 6580$~\AA), in order
to remove possible contamination from H~II regions.  We also obtained 
observations of the spectrophotometric standards LTT~377, LTT~1020, 
LTT~3218, LTT~4364, EG~21, Hiltner~600,
Feige~34, Feige~56, Feige~98, Kopff~27, BD~+25 3941 and BD~+33 2642
to determine our photometric
zero point \citep{stone1977,stone1983,mas1988}.  
A log of our observations is given in Table~\ref{table:log}.  At least one night of
each telescope run was photometric, which allowed us to put our PN
observations on the standard system.

The Magellan data were bias-subtracted at the telescope, 
using the IRAF\footnote{
IRAF is distributed by the National Optical Astronomy Observatory,
which are operated by the Association of Universities for Research
in Astronomy, Inc., under cooperative agreement with the National
Science Foundation.} reduction system.  The MagIC camera has a 
small (60--113 milliseconds), but known, shutter error that varies 
with position on the detector
\footnote{see http://www.ociw.edu/lco/magellan/instruments/MAGIC/shutter/index.html 
for a full discussion of this effect}.  For our target images, 
whose exposure times range from 200 to 1800s, the effect of the
shutter error is negligible.  However, for our dome-flat frames, and
standard star images, this effect could be significant.  These data 
were corrected for the shutter error by multiplying each 
image by a shutter correction image.  The shutter correction image was
constructed by dividing a long dome flat exposure of 20s in length by
an average of 20 one second exposures.  This process was repeated five
times so that a high signal-to-noise correction image was constructed.
After the flat-field exposures were corrected for the shutter error,
all of the data were flat-fielded using IRAF.  

The OPTIC data taken at WIYN were reduced following standard CCD mosaic 
imager procedures using the IRAF MSCRED package \citep{valdes2002}.
Because we did not use OPTIC's electronic tip/tilt compensation 
feature for these observations, no special processing steps were 
required beyond the usual overscan correction, bias subtraction, 
and flat-fielding. Unlike the MagIC camera, no significant shutter 
corrections were required.

The individual galaxy images were then aligned using the IMSHIFT 
task within IRAF.  The shifts were determined by measuring positions 
of stars common to all frames.  The images were then averaged using 
the IMCOMBINE task.  Finally, as the images were oversampled spatially 
($\approx$ 10--13 pixels FWHM for the Magellan data, and $\approx$ 4--8 
pixels FWHM for the WIYN data), the data were binned up into 
squares of 2 x 2 or 4 x 4 pixels.  This is known to 
improve the signal-to-noise of stellar detection and photometry 
\citep{harris1990}, as long as the binning is less than the critical 
sampling level.  The final [O~III] Magellan images of each galaxy are shown
in Figure~\ref{fig:images}, and the final WIYN image of NGC~4526 is 
shown in Figure~\ref{fig:n4526image}

\section{Searching for Planetary Nebulae candidates}

Planetary nebulae candidates were found on our frames using the 
semi-automated detection code of \citet{ipn2}.  Briefly, PN candidates 
should appear as point sources in the [O~III] $\lambda$ 5007 images, 
but should be completely absent in the off-band images, and
weak or absent in the H$\alpha$ image.  The code searches for 
objects that match these properties in two different ways: by 
using a color-magnitude diagram, and through a difference image 
analysis.  Both methods find many of the same objects, but there
is a population of candidates that can only be found in one method, and
not the other \citep{ipn2}.  This automated code has been rigorously tested,
with comparisons of previous manual searches of PN candidates, and by 
artificial star experiments (see Feldmeier \etal 2003 for full details).    
These previous tests have found that the automated detection code finds
virtually all candidates that have a signal-to-noise of nine or larger, and
also finds the vast majority of candidates below this limit. 

\begin{figure}
\figurenum{4}
\label{fig:examples}
\epsscale{0.65}\plotone{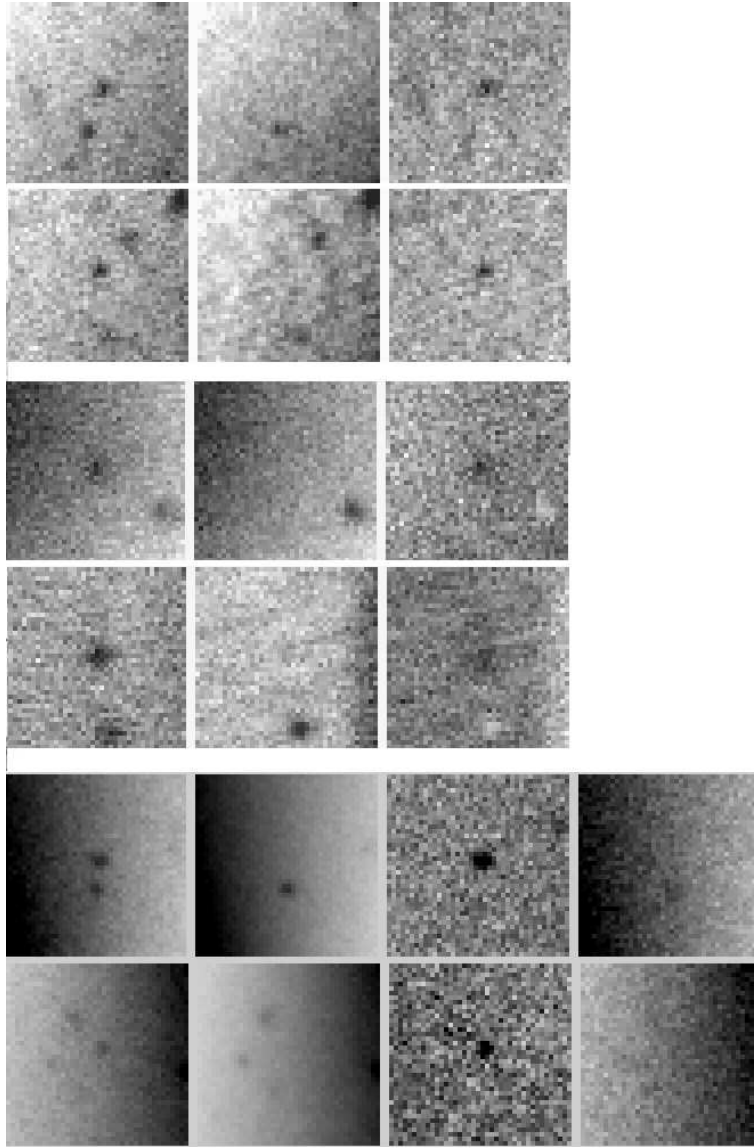}
\caption{Some examples of planetary nebula candidates in our target galaxies.  From
top to bottom, these candidates are from NGC~1316 (rows 1-2), NGC~1380 (rows 3-4), and
NGC~4526 (rows 5-6).  Column~1 (leftmost) shows a small region of the on-band image 
centered on each source, Column~2 shows the identical region in the off-band image,
and Column~3 shows the difference of the on-band and off-band images.  For NGC~4526, 
we have images in the H$\alpha$ band, and these are included in Column~4.
For each galaxy, we present one PN candidate near the bright luminosity function cutoff (upper), 
and the other candidate near the adopted photometric completeness limit (lower).  
All of these candidates have stellar image profiles, are present in the on-band 
and difference images, but are completely absent in the off-band image, and are 
weak to nonexistent in the H$\alpha$ images, making them likely PN candidates.}
\end{figure}

Once preliminary lists of candidates were compiled, they were
screened through a number of tests to ensure that they were 
genuine.  The tests included removing 
duplicate objects, removing candidates around saturated stars and other
bad regions of the image, removing any object that had a signal-to-noise 
less than the cutoff value of four \citep{ipn2}.  Since these images were 
relatively small, one of us (J.F.) also manually ``blinked'' the on-band and 
off-band images to ensure that no genuine candidates were missed 
by the automated detection code.  No such objects were found in the
Magellan data, but a few PN candidates were found close to the nucleus
of NGC~4526 that the detection code missed.  We believe this is due
to the steep and varying sky background from the galaxy, and not
due to a flaw in the detection code.
Finally, each planetary candidate was visually inspected, and any 
object that did not fit the selection criteria was removed.  In total, 
we identified 45 PN candidates in NGC~1316, 44 PN candidates
in NGC~1380, 94 PN candidates in NGC~4526, and no PN candidates in 
NGC~524 and NGC~1448.  A selection of PN candidates found in NGC~1316, NGC~1380
and NGC~4526 are displayed in Figure~\ref{fig:examples} for illustrative purposes.
For each galaxy, we display a candidate near the bright end of the 
luminosity function, and an object near the photometric completeness limit to
properly compare our detections as a function of signal-to-noise.

\section{Photometry and Astrometry of the PN candidates}

The PN candidates were measured photometrically using the IRAF version
of DAOPHOT (Stetson 1987), and flux calibrated 
using \cite{stone1977,stone1983,mas1988} standard stars and the procedures 
outlined by Jacoby, Quigley, \& Africano (1987).  The resulting 
monochromatic fluxes were then converted to $m_{5007}$ magnitudes using:
\begin{equation}
m_{5007} = -2.5 \log F_{5007} - 13.74
\end{equation}
where $F_{5007}$ is in units of ergs cm$^{-2}$ s$^{-1}$.
The scatter from the standard stars is small, 0.01 magnitudes for the
Magellan data, and 0.03 magnitudes for the WIYN data.

Equatorial coordinates were then obtained for each planetary 
nebula candidate by comparing its position to those of USNO-A 2.0
astrometric stars (Monet 1998; Monet \etal 1996) on the same
frame.  To calculate the plate coefficients, the FINDER astrometric 
package within IRAF was used.  Due to the small angular sizes
of the Magellan images, the final fit used approximately 5--10 USNO-A 2.0 
stars.  The WIYN image was better constrained, with 23 USNO-A 2.0 
stars in the final solution.  
The external errors from the USNO-A 2.0 catalog are believed 
to be less than 0.25 arcseconds (Monet 1998; Monet \etal 1996), and our fits
are generally consistent with this uncertainty.  All coordinates are 
J2000 epoch.  The magnitudes and coordinates for the PN candidates 
in NGC~1316, NGC~1380, and NGC~4526 are given in Tables 3--5.  The 
names of the candidates follow the new uniform naming convention
of extragalactic planetary nebulae (Jacoby \& Acker 2006).

\section{Fitting the PNLFs and Finding Distances} 
\subsection{NGC~1316 and NGC~1380}

\begin{figure}
\figurenum{5}
\label{fig:pnlf}
\epsscale{1.25}\plotone{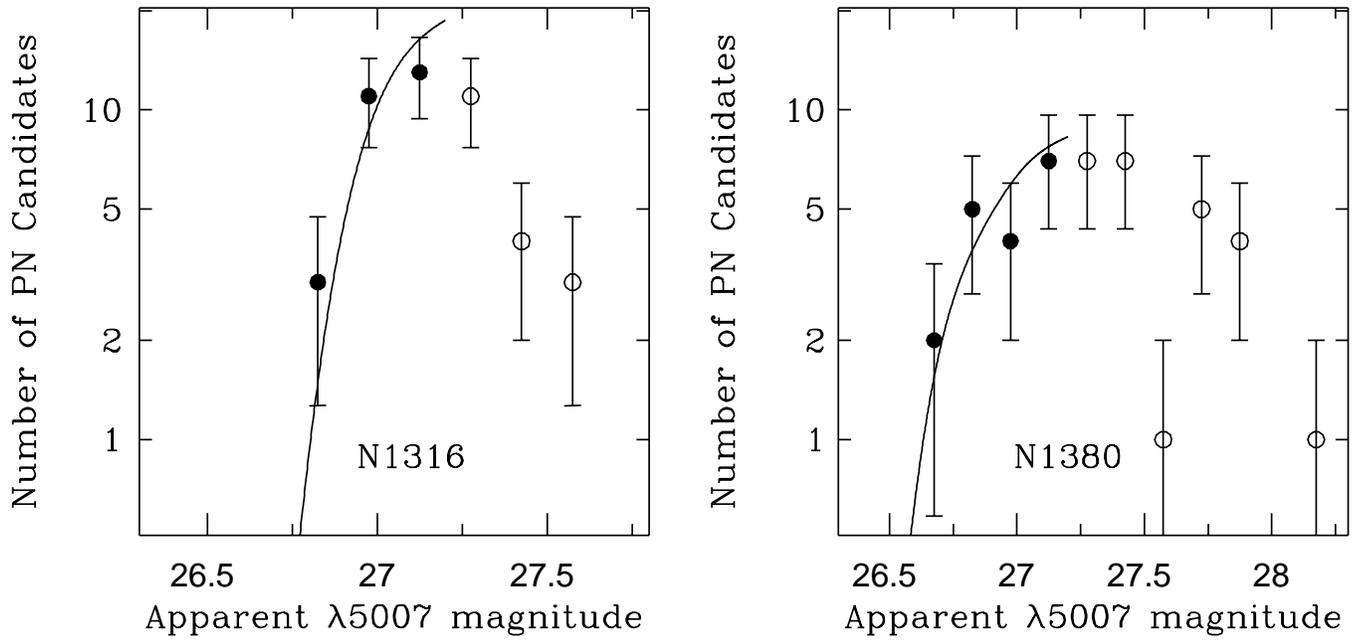}
\caption{The planetary nebula luminosity functions for NGC~1316 and
NGC~1380 binned into 0.2~mag intervals.  The solid lines 
represent the empirical PNLF of equation (2)
convolved with the mean photometric error vs.~magnitude relation 
and translated to the most likely distance modulus for each galaxy.  The
solid circles represent objects in our statistical PN samples; the open
circles indicate objects fainter than the completeness limit that were not 
included in the maximum likelihood solution.}  
\end{figure}

Figure~\ref{fig:pnlf} shows the observed PNLFs for both NGC~1316 and NGC~1380.
The luminosity functions follow a power law at faint magnitudes, but 
abruptly drop as a bright
limiting magnitude is reached.  This distinctive feature allows us to
obtain distances to galaxies at high precision.

In order to derive PNLF distances and their formal uncertainties, we followed
the procedure of \citet{pnlf2}.  We took the analytical form of the PNLF:
\begin{equation}
N(M) \propto e^{0.307 M} \{ 1 - e^{3 (M^* - M)} \}
\end{equation}
convolved it with the photometric error vs.~magnitude relation derived from 
the DAOPHOT output (given in Table~\ref{table:error}), and fit the resultant curve 
to the statistical
samples of PN via the method of maximum likelihood.
Since our ability to detect PN in these two galaxies was not a strong 
function of position, we determined the completeness for both galaxies
by noting where the PNLF (which should be exponentially increasing) 
begins to turn down.  This corresponds to an effective signal-to-noise
ration of approximately nine for our candidates.  To correct for
foreground extinction, we used the $100 \mu$ DIRBE/IRAS all-sky map of
\citet{schlegel1998}, and the reddening curve of \citet{ccm}, which 
corresponds to $A_{5007} = 3.56 E(B-V)$.  Finally, to
estimate the total uncertainties in our measurements, we convolved the formal
errors of the maximum-likelihood fits with the errors associated with the 
photometric zero points of the CCD frames (0.01~mag), the filter response 
curves (0.05~mag), the definition of the PNLF (0.03~mag), and the 
Galactic foreground extinction \citep[0.02~mag][]{schlegel1998}.  We adopt a 
PNLF cutoff of $M^* = -4.47 \pm 0.03$, consistent with the 
revision of the zero point by \cite{pnlf12}.  

Based on these assumptions, the most likely
distance modulus for NGC~1316 is  $(m-M)_0 = 31.26_{-0.12}^{+0.09}$,
corresponding to a distance of $17.9^{+0.8}_{-0.9}$~Mpc.  This distance
assumes a foreground extinction of E(B-V) = 0.021 \citep{schlegel1998}.
The most likely distance modulus for NGC~1380 is 
$(m-M)_0 = 31.04_{-0.15}^{+0.11}$, corresponding to a distance of 
$16.1^{+0.8}_{-1.1}$~Mpc, and assuming a foreground extinction of
E(B-V) = 0.017 \citep{schlegel1998}.

For NGC~1380, we can also calculate the observed production rate
of planetary nebulae as a function of bolometric magnitude, which
is parameterized by the quantity $\alpha_{2.5}$ (Ciardullo 1995).  
To do this, we first determined
the stellar bolometric luminosity of the galaxy in our survey frame.
We fitted the galaxy light profile
of the off-band image using the ELLIPSE task in IRAF/STSDAS
(Busko 1996; based on the algorithms of Jedrzejewski 1987),
and compared the luminosity found to the total $V$ band magnitude
found by \citet{rc3}.  After accounting for the fraction of the galaxy
that could not be surveyed due to crowding, we 
find an apparent bolometric magnitude of 10.3.  
The corresponding $\alpha_{2.5}$ parameter, summed over
all distances was $9.1^{+6.8}_{-2.9} \times 10^{-9}$~PN-$L_{\odot}^{-1}$.
This value is within the expected range found for elliptical
galaxies \citep{rbciau}.

\subsection{NGC~4526}

\begin{figure}
\figurenum{6}
\label{fig:n4526pnlf}
\plotone{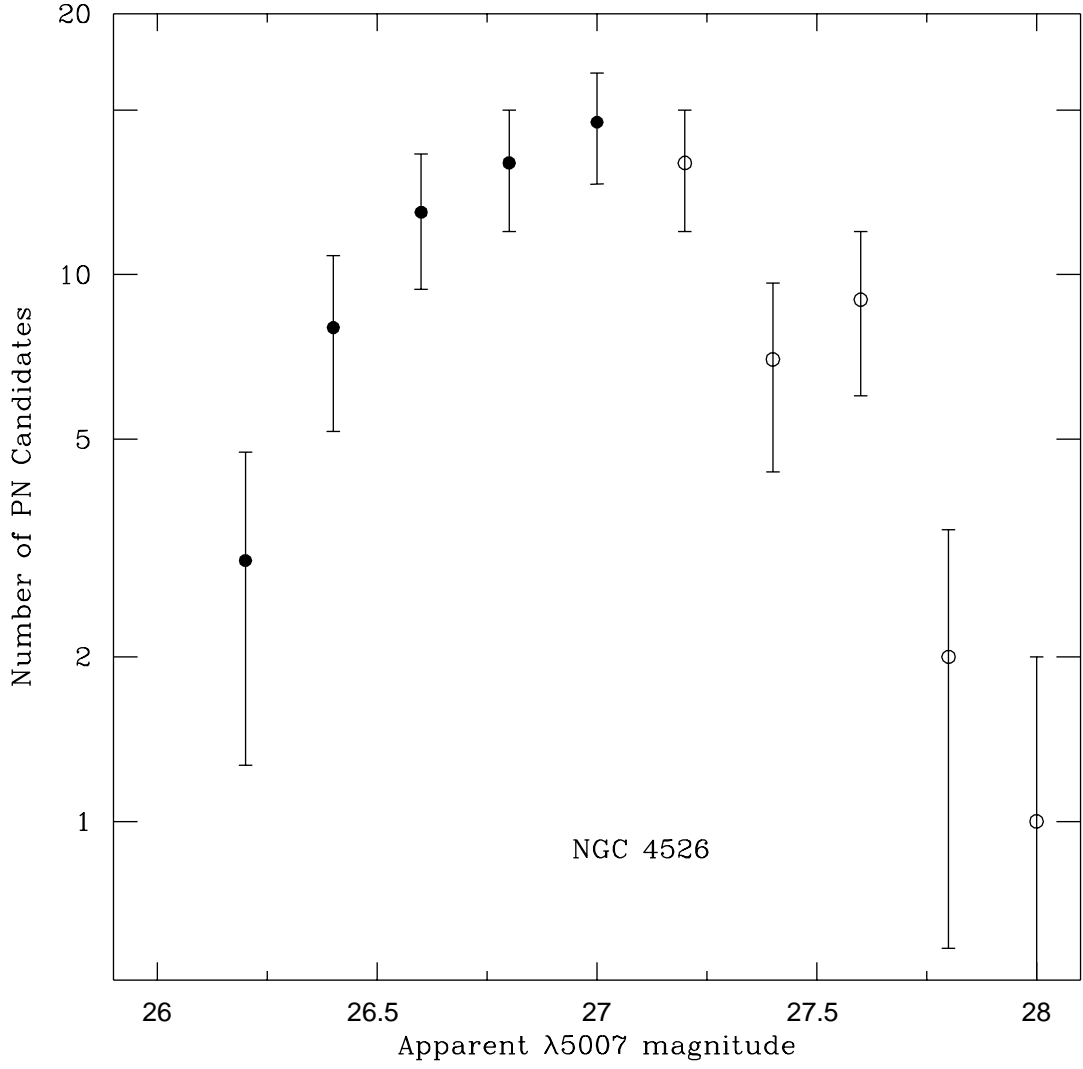}
\caption{The planetary nebula luminosity functions for NGC~4526 
binned into 0.2~mag intervals. The  solid circles represent objects in 
the statistical PN sample; the open
circles indicate objects fainter than the completeness limit that were not 
included in the maximum likelihood solution.  Note the shallow drop-off
in this PNLF compared to our other observations.  }  
\end{figure}

For NGC~4526, we plot the observed PNLF in Figure~\ref{fig:n4526pnlf}.
A visual inspection of the luminosity function shows that
it is declining slowly at the bright end, unlike the sharp cutoffs 
seen in the PNLFs of NGC~1316 and NGC~1380 and most other galaxies
in which the PNLF has been measured.  A more striking effect can 
be seen when we divide the NGC~4526 candidate
PN sample into two equal sub-samples.  Using the on-band image, the 
ELLIPSE task in IRAF/STSDAS, and the procedures outlined in \cite{icl1}, we 
measured the isophotal parameters of NGC~4526, and determined a
radial coordinate $r$ for each PN candidate.  We define
$r$ as  the geometric mean of the semi-major and semi-minor axes at 
the position of each PN candidate: 
$r = \sqrt{ab}$.  From this radial coordinate, we then divided the
94 PN candidates of NGC~4526 evenly 
into an inner and outer sub-sample.  The classification
for each object is given in Table~5, column 5 as the letter I or O,
and the two sub-samples are shown visually in Figure~\ref{fig:n4526image}.

\begin{figure}
\figurenum{7}
\label{fig:inout}
\plotone{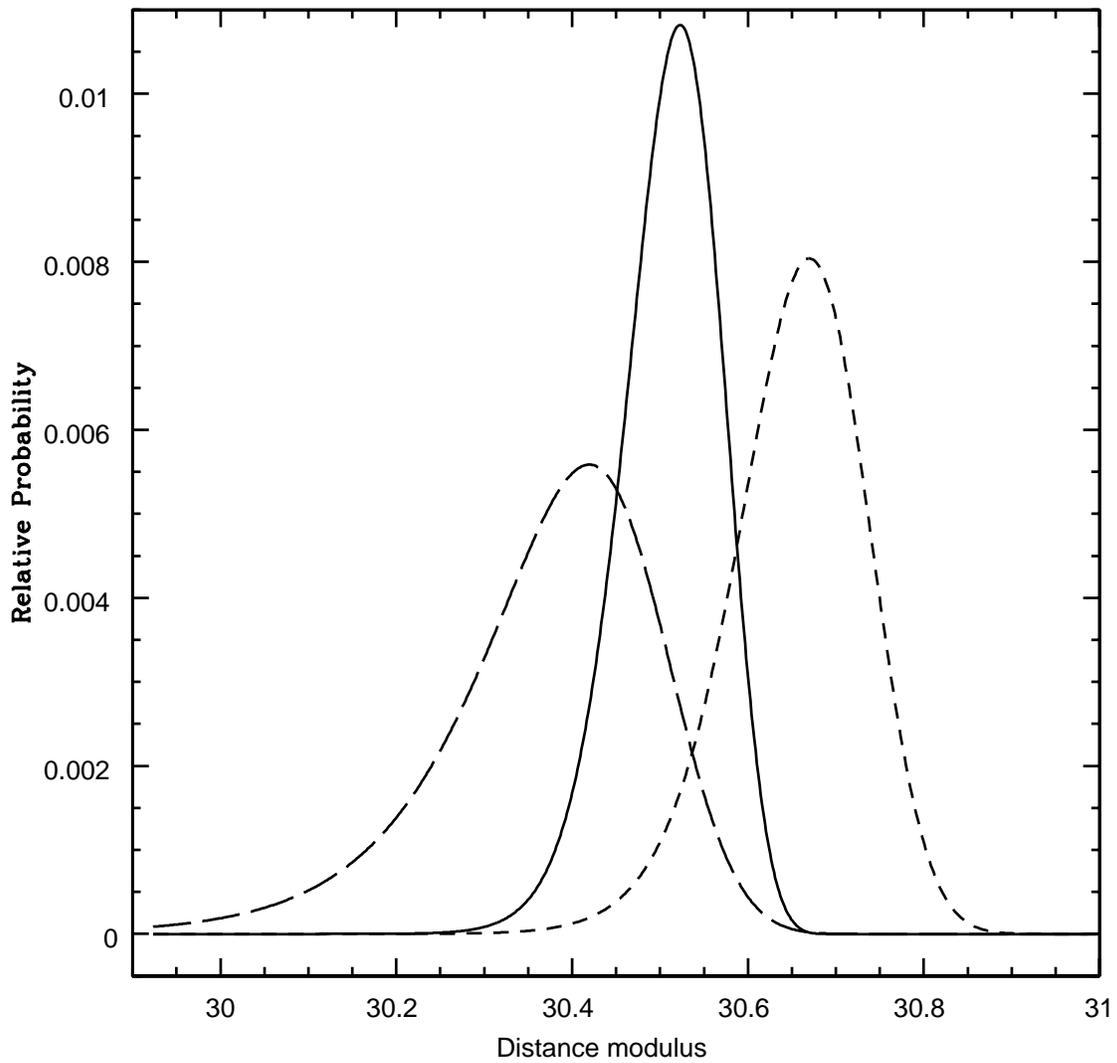}
\caption{The results from the maximum likelihood analysis for the
PNLF of NGC~4526.  The x-axis denotes the true distance modulus; the
y-axis is the probability that the observed PNLF is drawn from the 
empirical model at the given distance.  Corrections for extinction
and photometric error have been applied.  The solid line shows the
function for the entire data sample, the long-dashed line shows the
result for the 47 PN candidates in the outer sample, and the short-dashed line
shows the results for the 47 PN candidates in the inner sample.  There
is a clear offset in the derived distance between the different samples.}  
\end{figure}

Using the \citet{pnlf2} maximum likelihood method, we then found the PNLF
distances to NGC~4526 using the entire sample, the inner sub-sample, 
and the outer sub-sample.  Those results are plotted in Figure~\ref{fig:inout}.
There is a noticeable offset between the distance found from 
the three samples.  The entire sample has a best fitting distance
modulus of $(m-M)_0 = 30.51 \pm 0.05$.  In contrast, the outer sample 
has a best-fitting distance modulus of
$(m-M)_0 = 30.41^{+0.081}_{-0.141}$, and the inner sample
has a best-fitting distance modulus of
$(m-M)_0 = 30.66^{+0.066}_{-0.086}$, where the error bars
denote the results of the maximum likelihood fit only, and not any
other effects.  There is a difference of $\approx 2.1\sigma$ 
in distance modulus between the inner and outer sub-samples.

In PNLF observations of over 40 galaxies, these systematic behaviors 
with radius have only been seen previously in elliptical galaxies
in the Virgo cluster \citep{pnlf5}, and have been studied in detail in one
galaxy: M87 \citep{m87ipn}.  In that case, 
the flattening of the observed PNLF, and the differences between
inner and outer samples were found to be due to the presence of
intracluster planetary nebulae (IPN) in the Virgo cluster.  Of these
IPN, there will be some foreground to the galaxy of interest, which
will have a brighter apparent magnitude, and hence distort the
observed luminosity function.
The number of foreground PN detected in any region of our CCD field 
should be roughly proportional to the area of the field; in 
the case of NGC~4526, the outer sample contains $\approx$ 11.5 
times more area than the inner one, and hence is much more likely to 
contain contaminating objects.

\begin{figure}
\figurenum{8}
\label{fig:lfcomp}
\plotone{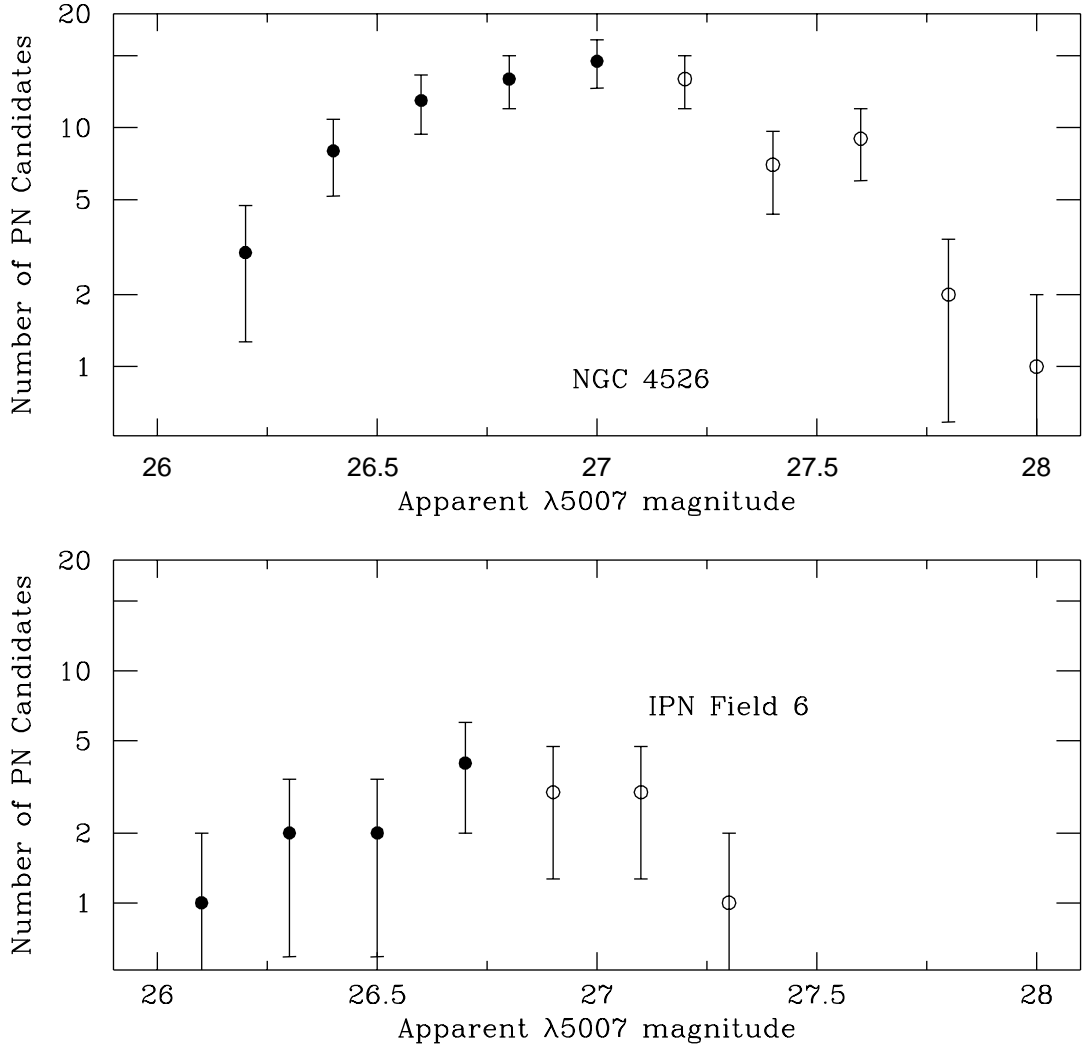}
\caption{A direct comparison of the observed PNLFs of NGC~4526, and from 
the intracluster planetary nebulae field IPN 6, which is located
$47\parcmin5$ ($\approx$ 210 kpc) away from NGC~4526.  As can be
clearly seen, the brightest IPN have comparable magnitudes to 
the brightest PN candidates in NGC~4526.
}  
\end{figure}

NGC~4526 is located in subclump ``B'' \citep{vc6} of the Virgo cluster,
several degrees to the south of M87, but there is abundant evidence
to suggest that IPN are the cause of the distance offsets observed.
Several hundred IPN candidates have been observed in multiple fields
of the Virgo cluster \citep[][and references therein]{ipn3,aguerri2005}.
In particular, Virgo IPN Field 6 \citep{ipn2} lies less than
$47\parcmin5$ 
from NGC~4526.  Assuming a mean Virgo distance of $\approx$ 15 Mpc, 
this corresponds to a linear transverse distance of only 
$\approx$ 210 kpc.  Figure~\ref{fig:lfcomp} directly compares the 
PNLF of NGC~4526 to that of IPN Field 6.  The luminosity functions 
sample comparable brightness, and several IPN candidates in 
Field~6 have apparent magnitudes similar to the brightest PN candidates 
observed in the NGC~4526 field.  If we scale the areas and the photometric
depths of the two different observed fields, we find that we would expect
at least two IPN objects within the magnitude range
of m$_{5007} = 26.1-26.7$ and within the angular area of the NGC~4526
inner and outer samples.  However, since there is known spatial structure
in the IPN distribution of Virgo \citep{ipn3,aguerri2005}, 
significant departures from this statistic are possible.

If we could separate the IPN distribution from the PN bound to the galaxy,
in principle we would obtain a distance as precise as any other PNLF 
observations.  However, without further information, it is problematic
to separate the two samples cleanly.  We therefore adopt the inner 
sample for our distance determination, as it should contain 
an order of magnitude less
foreground contamination.  In order to be conservative, we add 
a 0.15~mag systematic error in quadrature with our other errors to 
account for potential foreground contamination.
We then follow the identical procedures for finding the distances
given above.  We find the most likely distance modulus for NGC~4526 to be 
$(m-M)_0 = 30.66 \pm 0.2$, assuming a foreground extinction of
E(B-V) = 0.022 \citep{schlegel1998}.  This distance modulus 
corresponds to a distance of $13.6^{+1.3}_{-1.2}$~Mpc.  We should be able 
to improve our measurement of the distance by obtaining radial velocities 
of the PN candidates around NGC~4526.  It is unlikely that the foreground 
IPN will have the same radial velocities as PN bound to the galaxy, and
therefore we will be able to separate the two samples in a straightforward
fashion.

\section{Limits on Distance from Non-Detection}

In the case of NGC~524, and NGC~1448, we detected no planetary nebulae
candidates, so we can only obtain a lower limit to the distances in
these galaxies.  Normally, this is done by determining the 
completeness function as a function of magnitude 
through artificial star experiments, adopting a completeness limit, 
and finding the distance modulus that would correspond to that 
limit, assuming that limit corresponded to a PN with the 
brightest absolute magnitude, $M^*$.

Although the completeness function generally drops steeply
as a function of magnitude once a critical level has passed, 
it is not a simple step-function.  For most imaging surveys, there is 
also a transition zone, with a range of approximately one magnitude, 
where the completeness is declining, but is still non-zero.  We can 
obtain more accurate and robust distance limits by taking the full 
completeness function into account.  

Since this situation is generally applicable to any distance indicator
that uses a luminosity function, such as the Globular Cluster Luminosity
Function distance indicator (GCLF; see a recent review by Richtler 2003), 
or the Tip of the Red Giant Branch distance indicator (TRGB; see the
review by Freedman \& Madore 1998), we give a formal derivation 
for this procedure below.  We note that this formalism assumes
that the luminosity function of the distance indicator is well 
determined for the galaxy type in question.  If the luminosity function 
is not well determined, the formalism will be of limited usefulness.  

\subsection{Mathematical Formalism}

The number of objects expected, $N_{exp}$, in an imaging survey of 
a limited photometric depth can be expressed as follows:

\begin{equation}
N_{exp} = \frac{\int_{-\infty}^{+\infty} L(m) f(m) dm }
{\int_{-\infty}^{+\infty} f(m) dm }
\end{equation}
where $L(m)$ is the apparent magnitude luminosity function and
$f(m)$ is the completeness function, defined as the probability of
detecting an object with an apparent magnitude $m$.  For this analysis, 
we will assume 
that $f$ is solely a function of $m$, though this analysis can be 
expanded to include other effects, such as position or object surface
brightness.   We will also rewrite the luminosity function as follows:
\begin{equation}
L(m) = N_{0} \Phi(M + \mu )
\end{equation}
where $N_{0}$ is a normalization constant, $\mu$ is the distance modulus,
and $\Phi(M)$ is the absolute magnitude luminosity function.  
The actual number of objects detected, $N_{obs}$, is $N_{exp}$, 
subject to the Poisson distribution:

\begin{equation}
P(N_{obs}|N_{exp}) = \frac {N_{exp}^{N_{obs}}}{N_{obs}!} e^{-N_{exp}}
\end{equation}
We are interested in the case where no objects were detected
$P(0|N_{exp}) = e^{-N_{exp}}$ \hfil\linebreak 
or $P(0 | N_{0}, \mu, \Phi(M), f(m))$.  
In this case, what limits can we
place on $\mu$, given values of $N_{0}$, $\Phi(M)$ and $f(m)$?

Given a typical luminosity function that is rising to 
fainter absolute magnitudes, we can intuitively see that the distance limits 
will strongly depend on the total number of objects that are 
actually present, and hence the value
of $N_{0}$.  Given a pathological galaxy that contained none of the 
objects searched for in the imaging survey, there would be no effective limit 
to the distance.  Conversely, a galaxy with a very large number of 
targeted objects would have a distance limit that would be proportional 
to the inverse of the completeness
function 1 - $f(m)$.  Unfortunately, instead of a single distance 
limit, we now have a two dimensional curve in a space of $\mu$ and
$N_{0}$ with the general property that the smaller the value of $N_{0}$,
the smaller the distance modulus limit $\mu$. 

However, despair is unnecessary.  If we have some prior knowledge 
on the total number of objects we should expect in our survey region, 
we can constrain the values of $N_{0}$ substantially, and hence 
the distance limit.  Let us suppose that the number of objects 
present in the survey area is proportional to the 
galaxy luminosity, $L_{galaxy}$ that is present in that same area 
and in some photometric system, $x$: 
\begin{equation}
N_{0} = R L_{galaxy, x}
\end{equation}
Note that although we assume that the efficiency rate, $R$, can be expressed
as a simple constant, one could integrate a more complex rate over the 
entire survey area, and get similar results.  The above equation can 
be rewritten as follows:

\begin{equation}
N_{0} = R L_{\odot, x} 10^{-0.4 (m_{galaxy, x} - \mu - M_{\odot, x})}
\end{equation}
where $L_{\odot, x}$ and $M_{\odot, x}$ are the luminosity and 
absolute magnitude of the Sun in that same photometric system.  
If we can obtain the 
apparent magnitude of the galaxy from other observations 
and limit the efficiency rate over some range ($R_{min} < R < R_{max}$),
we can reduce the two-dimensional ($\mu, N_{0}$) curve to a 
curve segment whose arc-length is dependent on how small the range of $R$ 
can be constrained.  A robust distance modulus limit can then be adopted as
the minimum $\mu$ allowed.

\subsection{Application}

Given the formalism above, we now apply this method to our observations 
of NGC~524 and NGC~1448.  We now assume that the 
completeness function can be approximated 
in an interpolated form first given by \citet{fleming1995}:
\begin{equation}
f(m) = 1/2 \left[1 - \frac{\beta(m-m_{lim})}
{\sqrt{1+\beta^{2}(m-m_{lim})^{2}}}\right]
\end{equation}
where $m$ is the apparent magnitude, $m_{lim}$ is the magnitude where the 
photometric completeness level reaches 50\%, and the parameter
$\beta$ determines how quickly the completeness fraction declines around the
range of $m_{lim}$.  

\begin{figure}
\figurenum{9}
\label{fig:comp}
\plotone{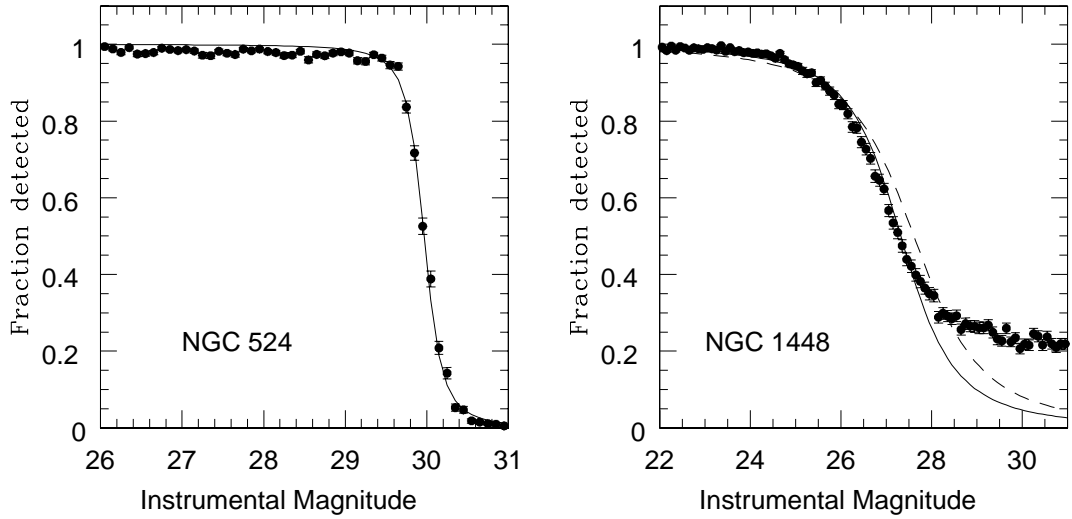}
\caption{The completeness function for NGC~524 and NGC~1448 as a function
of magnitude, with the error bars derived assuming the 
binomial distribution.  The solid line is the 
best-fitting function of \citet{fleming1995}.  In the case of
NGC~524, the fit is excellent, but due to crowding, the original fit of NGC~1448
(given as the dashed line) is relatively poor.  After re-fitting the function, 
omitting the points fainter than an instrumental magnitude of 28 which are
effected by crowding, we obtain a much better fit (given as the solid line).}  
\end{figure}

We determined these two parameters by adding 
artificial stars using the ADDSTAR task within DAOPHOT over a range 
of magnitudes to our data frames.  To keep crowding from artificial stars
from affecting our results, we added the stars in groups of 100.  We 
then searched the frame using the DAOFIND command within 
DAOPHOT, and found the recovery rate as a function of magnitude.  
We repeated this process until 
the completeness functions were well determined, at least 500 times in 
all.  The results of these experiments are  shown in 
Figure~\ref{fig:comp}, with the best fit to the function by 
\citet{fleming1995}.  

As can be clearly seen, the function fits well to 
the completeness simulations of NGC~524.  However, the \citet{fleming1995} function is
a poor fit to the completeness simulations of NGC~1448.  This is due to
crowding from H~II regions and other non-stellar sources in the galaxy
image, and can be seen as the $\approx$ 20\% false detection 
rate observed at faint magnitudes.  To provide a more realistic fit to the 
completeness results for NGC~1448, we re-fit the completeness function, 
but excluding all points fainter than an instrumental magnitude of 28.  
This revised fit is shown in Figure~\ref{fig:comp} as the dashed line, 
and we adopt for all further analyses.  The best-fitting values
for $\beta$ and $m_{lim}$ were 4.41 and 29.97 for NGC~524, and
0.800 and 27.32 for NGC~1448, respectively.  The values for $\beta$ and $m_{lim}$ 
are significantly different for each of the galaxies due to the environment of the
two images: the completeness function of NGC~524 is almost purely a case 
of photometric completeness, while NGC~1448's completeness function is 
a combination of photometric completeness and source confusion completeness. 
Nevertheless, the \citet{fleming1995} function is a reasonable fit to our artificial
star results for both galaxies.

With the completeness function established, we now place limits on
the efficiency rate of planetary nebulae production.  Historically, this 
has been parameterized by the value $\alpha_{2.5}$, which is the number 
of PN within 2.5 magnitudes of the peak magnitude, $M^{*}$, divided 
by the stellar bolometric luminosity.  Accordingly, we will denote 
the number of planetaries by $N_{2.5}$.  Assuming the PNLF:
\begin{equation}
N(M) = N_{0} e^{0.307 M} \{ 1 - e^{3 (M^* - M)} \}
\end{equation}
and integrating we find the following normalization for $N_{0}$, our adopted
normalization constant:
\begin{equation}
N_{2.5} = 3.38928 N_{0}
\end{equation}

Ciardullo (1995) found that in a sample 
of 23 elliptical galaxies, lenticular galaxies, and spiral bulges, the
$\alpha_{2.5}$ parameter ranged from 
$\alpha_{2.5} = 50 \times 10^{-9}$~PN-$L_{\odot}^{-1}$ to 
$\alpha_{2.5} = 6.5 \times 10^{-9}$~PN-$L_{\odot}^{-1}$ overall.  However,
\citet{rbcalpha} and \citet{rbciau} also found that 
the $\alpha_{2.5}$ varied systematically
with parameters such as the Mg$_{2}$ absorption line index, the ultraviolet
color of the galaxy, and the absolute magnitude of the galaxy.  For 
moderately luminous galaxies, similar to NGC~524, the
$\alpha_{2.5}$ parameter varied between 
$10-30 \times 10^{-9}$~PN-$L_{\odot}^{-1}$.  We adopt this range of
$\alpha_{2.5}$ for our distance limit calculation in this case.  
For NGC~1448, our image contains components from the disk, bulge, and
halo, and therefore our limit on the $\alpha_{2.5}$ parameter is
weaker.  We therefore adopt a larger range of $10-40 \times 10^{-9}$~PN-
$L_{\odot}^{-1}$ in this case.

Now, we must determine the apparent bolometric magnitude for each 
of our galaxy frames.  We determined these by first adopting the
total $V$ magnitudes from \citet{rc3}.  We next determined the
fraction of the total light from each galaxy that was included
in our data frames.  For NGC~524, we again used the ELLIPSE
task to determine the amount of light present in our survey
frame.  In the case of NGC~1448, we used the aperture photometry of
\citet{prug1998} to determine the fraction of light found in our
image.  We find a fractional value of 54\% for NGC~524, and 43\% for NGC~1448,
with an estimated error of a few percent in each case.  After
applying a bolometric correction of $-0.80$ (a value typical of older stellar
populations; Buzzoni 1989), and accounting for the area lost in our surveys
due to the high surface brightness regions in each galaxy, 
we find that the apparent bolometric magnitude in our images to be 
10.4 for NGC~524, and 10.3 for NGC~1448, with errors on order of 0.2
magnitudes.

\begin{figure}
\figurenum{10}
\label{fig:pnlimit}
\plotone{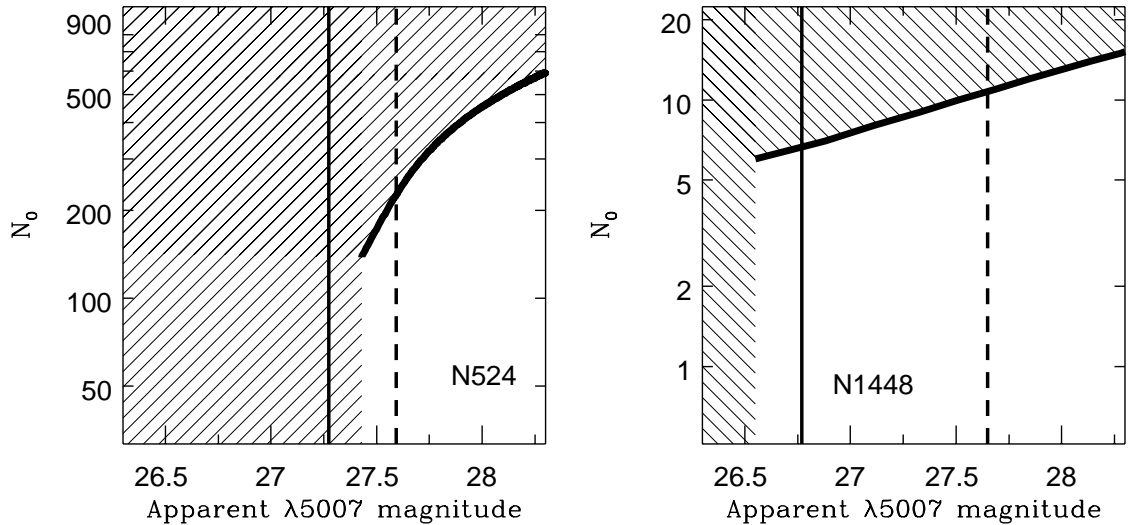}
\caption{The limits on the distance for NGC~524 and NGC~1448 as a function
of expected number of objects.  The thick solid line denotes the distance
limit, while the vertical solid and dashed line show the 90\% and 50\%
completeness levels from the artificial star experiments.  The filled
regions denotes where the distance limits are ruled out.  Note the
differences between the two curves, due to the differing completeness
functions.}  
\end{figure}

We now numerically calculate the distance limits by iterating over
values of $N_{0}$ and $\mu$, given the above constraints.  For each
value of $N_{0}$ and $\mu$, we find the probability of finding no
objects.  The best fitting curves for each galaxy are given in
Figure~\ref{fig:pnlimit}.  For NGC~524, the limiting apparent
[O~III] $\lambda$ 5007 magnitude, m$_{5007}$, ranges from  27.43 to 28.32.  
For NGC~1448, the m$_{5007}$ magnitude ranges from 26.55 to 29.81.
Assuming a foreground reddening of E(B-V) = 0.083 for NGC~524, and
E(B-V) = 0.014 \citep{schlegel1998}, we find that the lower limit
distance moduli for these galaxies to be $(m-M)_0 > 31.6$ (20.9 Mpc) 
for NGC~524 and $(m-M)_0 > 31.0$ (15.8 Mpc) for NGC~1448.    

\section{Comparison of Distance Results}

\subsection{Comparison to earlier PNLF results}
NGC~1316 was previously observed for planetary nebulae 
by \citet[][hereafter MCJ]{mcmillan1993}.  How do our new observations
compare with these earlier results?  MCJ observed 105 planetary nebulae
candidates in an $8\parcmin0$ by $8\parcmin0$ field under conditions
of $1\parcsec4$ seeing.  They determined a best-fitting PNLF distance 
of $(m-M)_0 = 31.12_{-0.15}^{+0.11}$, excluding any systematic errors, and
adopting the revised PNLF zero point.  This result is smaller 
than our result ($(m-M)_0 = 31.26_{-0.08}^{+0.05}$, excluding 
systematic errors).  The MCJ result assumed no foreground extinction 
to NGC~1316: if we adopt the \citet{schlegel1998} reddening, the 
difference in distance modulus increases to 0.21 $\pm$ 0.14 magnitudes.

Could this difference in distance be due to an error in photometry?
To directly compare our results, we searched for PN candidates common
to both data sets.  Due to the relatively small angular field of our 
Magellan data, compared to the wider field of MCJ, there are
only eight objects in the MCJ survey to compare against.  Of the 
eight objects found in this region by MCJ, we positively identified
four of them, which are listed in our catalog.  Of the remaining four, 
the automated code detected three of them, but they were removed as 
candidates in the screening process because they appeared 
too faint, or they had a non-stellar profile.
To compare the magnitudes directly, we included these omitted objects, 
and determined
m$_{5007}$ magnitudes for them in the identical manner as our PN candidates
The magnitudes from our study and MCJ are given in 
Table~\ref{table:compare}.  Although the scatter for each individual 
object is large, due to the faintness of the objects and velocity effects in the 
differing [O~III] $\lambda$ 5007 filters, the mean magnitude offset, weighted 
by the photometric errors, is consistent with zero.  Within the errors,
we cannot attribute the distance offset to photometry issues.

Instead, we attribute the distance offset to the possible inclusion
of background contaminating objects in the MCJ sample.  As we
have previously noted, three of the MCJ PN candidates in our
survey frame were rejected by our automated detection method as being
non-stellar.  This improvement is due to the difference 
in seeing: our new images are
almost a factor of two better in image quality.  Therefore, 
known contaminating objects such as Lyman-$\alpha$ sources 
that mimic the properties of faint PN (see Feldmeier \etal 2003 for
a discussion of these objects), are more likely to appear in the
MCJ sample than our sample.  Visual inspection of the MCJ 
luminosity function supports this hypothesis.  There is a noticeable 
excess of objects at the bright end of the PNLF of NGC~1316.
If these objects were background sources, and not genuine PN, then
the distance would be skewed slightly closer.  Spectroscopic
observations of the brightest PN candidates in our and MCJs sample
would verify the validity of this hypothesis.  For the moment,
due to the superior seeing and the smaller amount of contamination
expected, we adopt the newer determination as our best estimate
of the distance.

\subsection{Comparison with other distance indicators}

We now compare our distance measurements to those made by other authors.
For NGC~524 and NGC~1448, our distance limits of $(m-M)_0 > 31.6$ and
$(m-M)_0 > 31.0$, respectively are in good agreements with other
researchers.  \citet{jensen2003} obtained a distance modulus of   
$(m-M)_0 = 31.74 \pm 0.20$ for NGC~524 through a 
surface brightness fluctuation (SBF) measurement
and \citet{kris2003} obtained a 
distance modulus of $(m-M)_0 = 31.65 \pm 0.35$ for NGC~1448 through
the Tully-Fisher relation.  Our observations came tantalizing close
to these limits, especially in the case of NGC~524.
In the case of NGC~4526, \citet{dren1999} obtained a distance
to this galaxy using the GCLF distance indicator.  Their determined 
distance, $(m-M)_0 = 30.4 \pm 0.3$ although somewhat shorter than
ours, is in reasonable agreement, given the larger error bars.

However, our most fruitful comparison is with the SBF distance indicator.
\citet{jensen2003} reports the revised \citet{tonry2001} distances to  
NGC~1316 as $(m-M)_0 = 31.50 \pm 0.17$, NGC~1380 as 
$(m-M)_0 = 31.07 \pm 0.18$, and NGC~4526 as $(m-M)_0 = 30.98 \pm 0.2$.
Taking a weighted mean of the distance offsets to these galaxies, we find
that the SBF distances are $\Delta \mu = +0.18 \pm 0.13$ mag longer than
the our measurements.  We will discuss this offset in the next section, 
but here we note that NGC~4526 has a prominent dust lane, and 
\citet{pnlf12} noted that large distance offsets between SBF and 
PNLF are more common in galaxies with large dust lanes.  
  
\subsection{Global Comparison of distance scales}

We now combine our results from this work with other galaxies that have 
hosted SNe~Ia and have also determined 
PNLF distances.  These additional galaxies are 
NGC~5253 (SN 1972E), NGC~5128 (SN 1986G), NGC~3627 (SN 1989B), NGC~4374 
(SN 1991bg) and NGC~3368 (SN 1998bu), leading to a total of eight galaxies 
that can be studied.  We next compare these distances to those found 
for SNe~Ia host galaxies by the Cepheid and SBF distance indicators.

Table~\ref{table:distances} summarizes the Cepheid, SBF, and PNLF 
distances to galaxies that hosted SNe~Ia.  All Cepheid distances are 
on the Freedman \etal (2001) distance scale, and we have not applied 
any corrections for metallicity.  The SBF distances are from the 
compilation of \citet{tonry2001}, with the zero-point correction 
by \citet{jensen2003}.  For the purposes of obtaining high precision values, 
we have included only those SNe~Ia observed photoelectrically or with CCDs.  
Although extensive analysis of photographic observations of historical 
supernovae has been made, the difficulties of sky subtraction and 
transforming the photographic magnitudes to a modern photometric 
scale limit the precision of these observations \citep{bw1991,pierce1995,riess2005}.

From Table~\ref{table:distances}, we can directly compare global distance 
offsets between the three methods for the galaxies in question, and look for 
any systematic effects.  There are three host galaxies that are in common between 
the Cepheid distance indicator and the PNLF.  The weighted distance average, 
$(m-M)_{0, Cepheid} - (m-M)_{0, PNLF}$ is +0.04 $\pm$ 0.06 mag, in good agreement 
with results by \cite{pnlf12}.  There are also five host galaxies that are in
common between the SBF distance indicator and the PNLF.  In contrast, the weighted 
distance average between the SBF distance indicator and the PNLF 
is significantly offset, $(m-M)_{0, SBF} - (m-M)_{0, PNLF}$ = 
+0.22 $\pm$ 0.08 mag.  This would be expected from the analysis of the individual 
galaxies in \S 7.2.  Again, this offset is in approximate agreement with the results of 
\citet{pnlf12}, who found a difference of +0.30 $\pm$ 0.05 for larger sample of 
28 galaxies with both SBF and PNLF distances.  However, the SBF distance moduli in 
\citet{pnlf12} must by decreased by 0.12 mag to be consistent with the revised  
\citet{jensen2003} SBF distance moduli, which are on the revised \citet{freedman2001} 
zero point.  The \citet{pnlf12} difference in distance moduli therefore 
becomes +0.18 $\pm$ 0.05, which agrees better with the observed difference between the 
smaller sample of SNe~Ia host galaxies.  \citet{pnlf12} suggested that the 
distance offset between the SBF and PNLF distance scales is due to a small 
amount of uncorrected reddening.  We will return to this point later in our 
discussion, but for the moment we will adopt the distances in 
Table~\ref{table:distances} for comparison purposes.

\section{Measuring the zero point of SN~Ia and estimating the Hubble Constant}

With the distances to the host galaxies adopted, we now turn to determining the 
absolute magnitudes of SNe~Ia, and finding updated estimates for the Hubble constant
with zero point calibrations given from each of the three distance indicators 
discussed in \S7.3.  Since we are interested in searching for any systematic 
differences in SNe~Ia properties as a function of stellar population, 
comparing the Hubble constants derived from the Cepheid, SBF, and PNLF distance 
scales may give insights into the importance of such population effects.  We perform 
the zero point calibration twice: once for classical optical (BVI) and once for 
near-infrared (JHK) observations.  By comparing the Hubble constants derived from 
these different wavebands, we can also search for additional systematic effects, such
as abnormal reddening, or improper corrections for the known maximum 
magnitude-decline rate relation.

\subsection{Optical Calibrations}

We proceed as follows: first, taking advantage of the known color evolution of 
SNe~Ia from 30--90 days \citep{lira1995,phillips1999}, and using the methodology of 
\citet{phillips1999}, we estimate the host galaxy extinction.  With the extinction
determined, the absolute magnitudes of the SNe~Ia are now determined, and are given
in Table~\ref{table:absmag}.  Next, we correct the BVI light curves for the decline rate versus 
peak luminosity relationship for SNe~Ia.  Specifically, we compare the peak 
magnitude of the light curve versus the light-curve decline rate 
parameter $\Delta m_{15}(B)$.  For the nearby calibrators, we adopt the ideal 
criteria of Riess et al. (2005).  These are: 1) photometry from photoelectic 
(PE) observations or CCDs, 2) low host galaxy extinction (A$_{V}$ $<$ 0.5 mag), 
and 3) observed before maximum.  The supernovae that meet these criteria are 
identified in the last column of Table~\ref{table:distances} as ``opt''.  
All other SNe~Ia are omitted from further analysis.  We recognize that these
criteria throw out even well-observed supernovae such as SN~1998bu,
whose high optical extinction rules it out of our sample (Suntzeff \etal 1999; Jha
\etal 1999).  However, we concur with Riess \etal (2005) that the path
to a more accurate Hubble constant is in utilizing the supernovae that
have the fewest potential difficulties.

To compare these nearby supernovae against a more distant sample, which
are in the quiet Hubble flow but not significantly affected by cosmic acceleration, 
we used the subset of 38 SNe~Ia with moderate redshifts (7000 \kms $<$ cz $<$ 24000 \kms) 
from the sample of \citet{riess2004}, known as the ``gold'' sample.  The gold 
sample SNe were selected because they do not suffer from any of the following: 1) an 
uncertain classification, 2) incomplete photometric record (e.g., poor 
sampling or color information, or non-CCD/PE observations), or 3) 
large host extinction (A$_{V}$ $>$ 1 mag).

\begin{figure}
\figurenum{11}
\label{fig:gold}
\epsscale{0.8}
\plotone{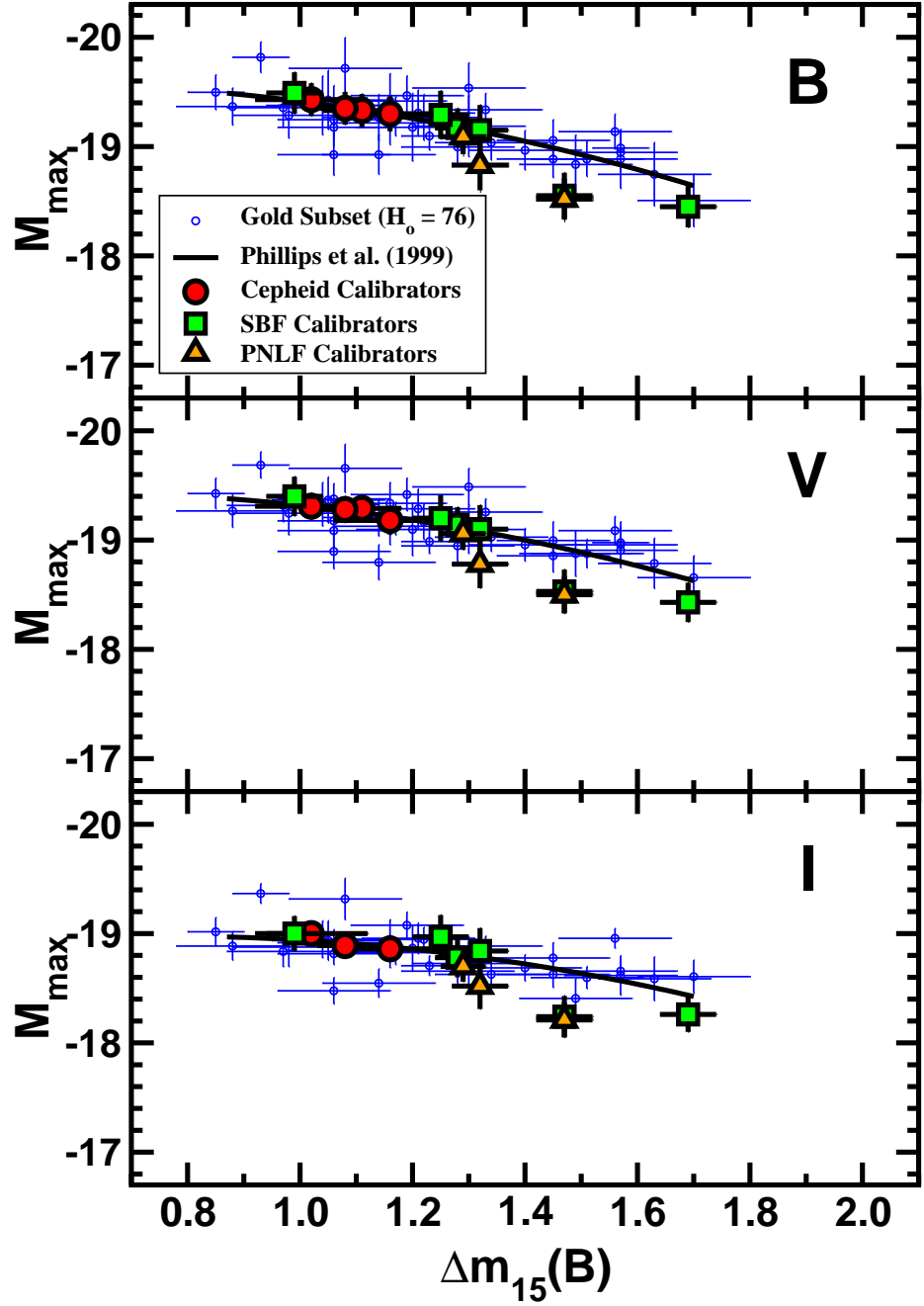}
\caption{The absolute magnitudes of SNe~Ia at maximum light in the $B$, $V$, and $I$ bands,
derived from the Cepheid, SBF, and PNLF distance indicators, and compared to
the ``gold'' subset of \citet{riess2004}.  The supernova maximum magnitude versus
\dm relation of \citet{phillips1999} is overplotted for reference.  
The point at \dm = 1.47 is SN~1992A, which is discussed in the text.}
\end{figure}

The results for Cepheid, SBF, and PNLF calibrators are summarized in 
Table~\ref{table:hubblebvi},
and are illustrated graphically in Figure~\ref{fig:gold}.  We note
that the errors given in both Table~\ref{table:hubblebvi} and Figure~\ref{fig:gold} are the
internal errors only.  These include errors in estimates of peak magnitudes, 
decline rates, host galaxy reddening, K corrections, and the fits to the luminosity 
versus decline rate relations.  An intrinsic dispersion of $\approx$ 0.16 mag in final 
luminosity-corrected peak magnitudes is assumed.  However, none of the systematic 
errors listed in Table 14 of \citet{freedman2001} are considered in this analysis.
From these results, we find that the results for the Cepheid and SBF calibrators are
 highly consistent, with a mean offset of only 1\%.  However, the  PNLF calibrators give a 
Hubble constant which is ~11 $\pm$ 7\% (internal errors only) greater than the other 
two methods, which is marginally discrepant.  This offset may be due, in part, to 
small number statistics.  For example, if we had used only the PNLF calibrators 
(1980N, 1992A, 1994D) to calibrate the SBF value for the Hubble constant, 
we would have derived a value of $\approx 79$ for this method, or only 5\% larger than 
the Cepheid based result.  

The difference of $\approx$ 0.32 $\pm$ 0.28 mag between SBF and PNLF distance 
moduli for NGC~4526 might imply that PNLF distance has not been fully corrected 
for intracluster PNLF (\S5.2).  However, a similar difference exists for another 
SN host, NGC 5128, which is not in a cluster environment, and 
therefore is unlikely to suffer from the same effect.  In any case, the conservative error 
bars we previously adopted for this galaxy's distance span the possible range of PNLF
values for this galaxy.  Interestingly, both the SBF and PNLF distances imply a 
low luminosity for SN 1992A, which falls $\sim$ 0.4 mag below the predicted luminosity 
for SNe~Ia with similar decline rates.  This is in contradiction with the results
of \citet{dren1999}, who argued that this supernova had a normal luminosity at
peak magnitude.  However, given the large error bars on all three direct 
distance indicators for this galaxy, this result is not definitive.

\subsection{Near-Infrared Calibrations}

In addition to the optical analysis, we also performed a separate analysis of 
SNe~Ia that have near-infrared observations (NIR) in the JHK bands.  
There are a number of
potential advantages to measuring the Hubble constant using near-infrared 
observations of SNe~Ia.  First, the effects of extinction are dramatically 
reduced in these bands, by up to a factor of $\sim 8$ \citep{ccm}, and there
is less sensitivity to unusual reddening laws \citep{kris2000}.  This
allows us to include supernovae that were rejected in the optical
calibration, such as SN 1998bu.  Second, it 
appears that SNe~Ia are well-behaved in the NIR, in that they obey the simple ``stretch''
model at maximum light \citep{kris2004}.  Third, and most importantly, observations 
to date show that SNe~Ia are excellent {\it constant} standard candles in the near-infrared 
\citep[][and references therein]{kris2004}, with a dispersion less than 0.20 mag.  

To calibrate the zero point of SNe~Ia in these bands, we follow a similar 
path as our previous analysis.  For this calibration, we have used the methodology of 
\citet{kris2004} for the SNe light curve fits.  In particular, the estimations of the 
time of maximum light are derived from BVI light curves, since the NIR observations
of earlier SNe~Ia were in general more sparsely sampled.  No luminosity corrections 
have been applied to the NIR magnitudes for decline rate differences. 

For the distant sample in the Hubble flow, we use the four SNe~Ia with JHK light curves from 
\citet{kris2004}, plus one SNe~Ia from \citet{kris2006} that have radial velocities 
between 7000 km/s $<$ cz $<$ 24000 km/s, and which were selected to not suffer from 
any of the following: 1) an uncertain classification, 2) incomplete photometric record 
(e.g., poor sampling or color information, or non-CCD/PE observations) in BVI, or 
3) large host extinction (A$_{H}$ $>$ 0.5 mag).  For the nearby calibrators, we adopt the 
following selection criteria: 1) low host extinction (A$_{H}$ $<$ 0.5 mag), 
2) observed before maximum in the optical, and 3) "normal" classification.  
SNe meeting these criteria are identified in the last column of 
Table~\ref{table:distances} as "NIR".

\begin{figure}
\figurenum{12}
\label{fig:jhk}
\epsscale{0.8}
\plotone{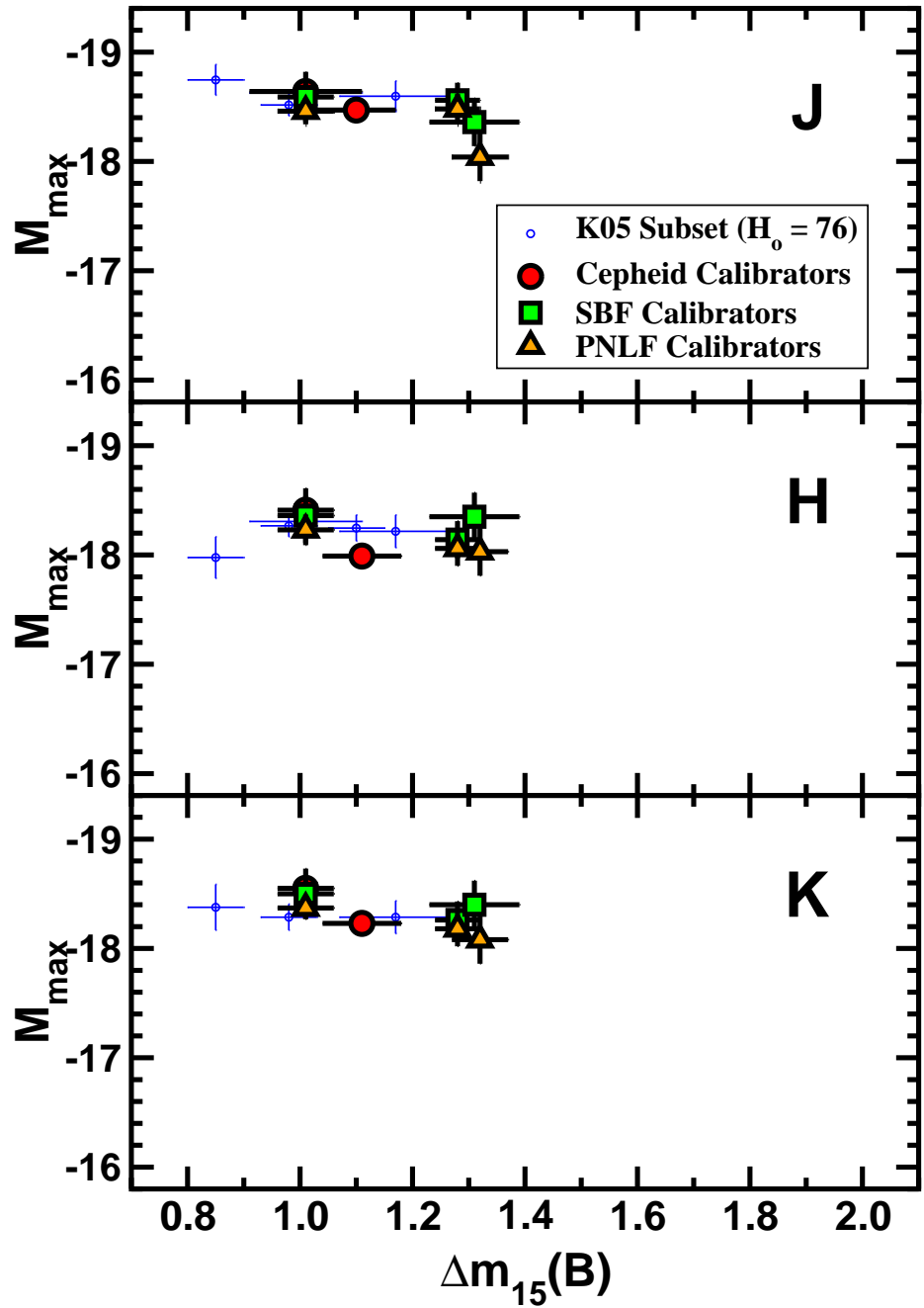}
\caption{The absolute magnitudes of SNe~Ia at maximum light in the $J$, $H$, and $K$
bands, similar to Figure~\ref{fig:gold}.  Note the lack of a strong correlation
with decline rate, as was originally found by \citet{kris2004}.}  
\end{figure}

The results for Cepheid, SBF, and PNLF calibrators are summarized in 
Table~\ref{table:hubblejhk},  and are illustrated graphically in Figure~\ref{fig:jhk}.  
Again, it should be noted that these errors are internal only.  These include errors 
in estimates of peak magnitudes, host galaxy reddening, and K corrections.  
An intrinsic dispersion of $\approx$ 0.16 mag \citep{kris2004} in final peak magnitudes 
is assumed.  The results for Cepheid and SBF calibrators are, again, quite consistent 
within the errors.  The calibration from the PNLF gives a Hubble constant which is
6 $\pm$ 10\% greater than the other two methods, well within the internal error bars.  
The agreement between the three distance scales is noticeably better than obtained 
using BVI light curves, perhaps because SN 1992A is not one of the NIR calibrators.

From these results, using near-infrared observations of SNe~Ia to measure the Hubble constant 
is clearly promising, but is currently limited by the small number of SNe Ia 
in the Hubble flow which have measured JHK light curves.  In particular, only a small 
range of light curve decline rates (0.97 $<=$ \dm $<=$ 1.16) are covered by the 
present sample, and this may lead to an underestimate of any potential systematic effects.
More observations will be needed to fully take advantage of this new method, and place
it on a sounder footing.

\section{Discussion}

After combining the results from both the optical and near-infrared calibrations, 
we find that the Hubble constants derived from SNe~Ia gives H$_{0}$ = 75 $\pm$ 3 \ho for 
Cepheid calibrators, H$_{0}$ = 76 $\pm$ 2 \ho for the SBF, and H$_{0}$ = 82 $\pm$ 3 \ho for the 
PNLF.  The difference between the PNLF and Cepheid values of H$_{0}$ is at the level
of 1.67$\sigma$, with the error bars almost overlapping.  This difference is at least 
partly a consequence of small number statistics.  Specifically, of the three 
PNLF SNe~Ia calibrators, one (SN 1992A) appears to 
be genuinely sub-luminous for its decline rate.  If we were to arbitrarily
eliminate this SN from our Hubble constant calculation, the BVI H$_{0}$ 
estimate from PNLF would decrease by ~3\%, but it would then be based on only 
two SNe.  More calibrators with PNLF distances are clearly needed to say 
something more definitive.

The difference between the PNLF and SBF values of H$_{0}$ is also of fairly 
low significance.  In this case, however, we might have expected a difference 
based on the fact that SBF distance moduli are, on average, 0.18 $\pm$ 0.05 
larger than PNLF moduli (\S 7.3), even when both are put on the Freedman et al. 
(2001) scale.  The PNLF and SBF methods react in opposite directions to reddening, and 
even a small amount of unaccounted internal extinction in the bulges of the 
calibrating spirals can lead to a large discrepancy in the derived distances.  
If both techniques are affected, then 
$\sigma_{\Delta \mu} = 7 \, \sigma_{E(B-V)}$ \citep{pnlf12}.  To account for the 
$\approx$ 5\% difference in the Hubble constants derived via the SBF and PNLF methods 
would require E(B-V) $<$ 0.02 of unaccounted internal reddening.  Given
the $\approx 16$\% uncertainty in determining Galactic foreground 
reddening \citep{schlegel1998}, and the $\approx 10$\% uncertainty 
in determining extinction in large-scale dust features of elliptical 
galaxies \citep{goud1994}, this small amount of internal extinction 
is quite plausible.  

However, thus far we have only discussed internal errors between the three
distance indicators.  As has been well known for decades, the external systematic
errors often dominate the uncertainties in the Hubble constant.  We now
briefly investigate some of the more recent systematic adjustments to 
the extragalactic distance scale relevant to SNe~Ia.  In particular, \citet{riess2005} 
have provided an updated calibration of the Hubble constant using SNe Ia 
with host galaxy Cepheid distances.  Besides adding two new calibrators 
(SN 1994ae and SN 1998aq), the primary differences between the \citet{riess2005} 
and the \citet{freedman2001} calibrations are: 1) application of a metallicity 
correction to the Cepheid distance scale and 2) an improved Period-Luminosity (P-L) 
relation, 3) elimination of poorly observed, highly reddened, and/or peculiar SNe.  
\citet{riess2005} apply a metallicity correction to the measured Cepheid distance 
moduli of -0.24 mag/dex, whereas the \citet{freedman2001} distance moduli do not 
include a metallicity correction.  In the case of the P-L relation, \citet{freedman2001} 
use the OGLE result for the LMC, whereas \citet{riess2005} use the OGLE relation 
truncated at periods below 10 days.

These changes, if correct, can have a significant effect on the Hubble constant derived from 
SNe~Ia.  \citet{riess2005} estimate the effect of invoking the metallicity correction 
to be a 4\% decrease in H$_{0}$.  To investigate the effects of this new calibration,
we looked at the effects on the absolute magnitude cutoff of the PNLF, $M^*$, for the 
PNLF relation assuming the same Cepheid calibrators used by \citet{pnlf12}, but 
correcting these for metallicity as per \citet{riess2005}.  If we take these corrections at
face value, we find that $M^*$ became brighter by up to 0.10 mag, implying a Hubble 
constant that is 5\% smaller, in good agreement with the \citet{riess2005} estimate.
The second effect of the different P-L relation is estimated by \citet{riess2005} 
to lead to a 2\% decrease in H$_{0}$.  To make this correction rigorously, we would have 
to independently re-fit the P-L relations for the galaxies that provide the Cepheid 
calibration for the PNLF.  Since this appears to be a small effect, we will simply 
adopt the \citet{riess2005} estimate and apply it to the PNLF distances as well.

Adopting these two modifications to our distance scales give 
H$_{0}$ = 72 $\pm$ 3 \ho for the Cepheid calibrators, 
H$_{0}$ = 72 $\pm$ 2 \ho for the SBF, and H$_{0}$ =  77 $\pm$ 3 \ho for the PNLF 
(internal errors only).  The Cepheid results compares well with the \citet{riess2005} 
value of 73 $\pm$ 4\ho.  This is as expected, since we used the same four calibrators 
and assumed the same distance moduli.  The small difference is due to differences 
in the methods used for calculating the distances of SNe Ia from their light curves.

We now estimate the systematic errors in these three measurements.  For the
Cepheid distance scale, we apply the three systematic effects adopted by 
\citet{riess2005}: 1) the LMC distance modulus uncertainty (0.10 mag), 2) the
error on the slope of the revised P-L relation (0.05 mag), and 3) the systematic
error of the fit to the ridgeline of the Gold sample (0.025 mag).  Therefore, the
Cepheid distance scale gives a systematic error of 0.115 mag, or about 5\% in the
value of H$_{0}$.  For the PNLF distance indicator, we must add in another
systematic term in quadrature, the observational error in determining M$^*$ 
(0.03 mag), and for the SBF an estimate of the errors in the SBF slope 
(about 0.03 mag).  However, when these are combined in quadrature with the   
previous systematic effects, they are effectively negligible.  The final results
become  H$_{0}$ = 72 $\pm$ 3 (random) $\pm$ 5 (systematic) \ho for the Cepheid 
calibrators, H$_{0}$ = 72 $\pm$ 2 (random) $\pm$ 5 (systematic) \ho for the 
SBF, and H$_{0}$ =  77 $\pm$ 3 (random) $\pm$ 5 (systematic) \ho for the PNLF.  If
we were to arbitrarily remove SN~1992A from our list, the PNLF value would become
H$_{0}$ =  74 $\pm$ 5 (random) $\pm$ 5 (systematic) \ho. 

If we take these estimated values of the Hubble constant, and compare them with the 
three year results from the Wilkinson Microwave Anisotropy Probe (WMAP), we
find a reassuring agreement.  Assuming a power-law flat $\Lambda$CDM model,
Spergel \etal (2006) finds a value for H$_{0}$ = $73.4^{+2.8}_{-3.8}$ \ho,
in good agreement with our estimations.  However, this agreement does depend on a 
number of assumptions, namely the assumption of a simple flat $\Lambda$ 
CDM model.  Non-standard cosmologies such as positively curved models without
a cosmological constant are consistent with WMAP results and can have Hubble
constants as low as H$_{0}$ = $30$ \ho (Spergel \etal 2006).  However,
these non-standard cosmologies are disfavored for a number of reasons, including the 
constraints from high redshift SNe~Ia results \citep{riess1998,perlmutter1999}, that
strongly imply a flat universe.  More realistically, we can compare the WMAP three-year results 
with joint constraints from other observations or assuming inflationary models.  
These differing adopted constraints give a rough estimate of the systematic
scatter that could be present in the WMAP results.  These estimates range from
H$_{0}$ = $68.7^{+1.6}_{-2.4}$ \ho to H$_{0}$ = $79.2^{+3.6}_{-6.8}$ \ho, though most
constraints have a much smaller scatter (70 $<$ $H_{0}$ $<$ 74 \ho; Spergel \etal 2006).  

In conclusion, from the evidence to date, it appears that absolute magnitudes of 
SNe~Ia appear to be broadly consistent, over a range of decline rates.  If we were 
to assume that the differences in our derived Hubble constants were solely due to some 
systematic effect of SNe~Ia, it would imply a difference of no more than $\approx 7\%$
in luminosity.  This is unlikely, as there may be additional systematic 
effects in the Cepheid, SBF, and PNLF distance scales at the few percent level.  
However, the case of SN~1992A, which appears to be $\sim$ 0.4 magnitudes fainter 
than most other SNe~Ia at the same decline rate, may be a signal for additional 
complexity in the absolute magnitudes of SNe~Ia.  We stress that these 
results are still tentative, due to the small number of calibrators, and the 
limited overlap between the Cepheid, SBF, and PNLF distance indicators.  As 
the number of modern observations of Type~Ia SNe increases, and additional 
distances are derived, these results should improve considerably. 

There are a number of observations that can improve the results presented
in this paper.  Images of our PN candidates in excellent seeing should remove
the vast majority of Lyman-$\alpha$ galaxy contaminants, as Lyman-$\alpha$ galaxies
at this redshift can be resolved in ground-based images (\eg~Hickey \etal 2004).
However, the most beneficial follow-up observations would be spectroscopy  
of the brightest PN candidates in NGC~1316 and NGC~4526.  These observations would 
confirm or deny the presence of contaminating objects in a straightforward
fashion.  In the case of NGC~1316, spectroscopic observations would clearly distinguish
between Lyman-$\alpha$ galaxies and genuine plantetaries by the 
width of the spectral lines, and the presence or absence of the [O~III] $\lambda 4959$ 
spectral line, which should be present in all genuine
planetaries.  In the case of NGC~4526, it is extremely unlikely that intracluster 
planetaries will follow the rotation curve expected for genuine galaxy PNe.
Ultimately, however, more high-precision distances to galaxies that host high-quality
supernovae will allow us to determine the Hubble Constant to the precision 
expected.

\acknowledgments

We thank Patrick Durrell and Lucas Macri for helpful discussions and we thank 
the Magellan and WIYN staff for their assistance with the observations.  
We also thank an anonymous referee for several suggestions that improved 
the quality of this paper.  

This work is supported by NSF through grant AST-0302030 (JF).
This research has made use of the NASA/IPAC Extragalactic Database 
(NED) which is operated by the Jet Propulsion Laboratory, California 
Institute of Technology, under contract with the National Aeronautics 
and Space Administration.  This research has also made use of the 
USNOFS Image and Catalogue Archive operated by the United States 
Naval Observatory, Flagstaff Station
(http://www.nofs.navy.mil/data/fchpix/).

{\it Facilities:} \facility{Magellan:Clay (MagIC)}, \facility{WIYN (OPTIC)}

\pagebreak

\clearpage

\begin{deluxetable}{llcll}
\tablenum{1}
\tabletypesize{\scriptsize}
\tablewidth{0pt}
\tablecaption{Target Information\label{table:target}}
\tablehead{
\colhead{Galaxy}
&\colhead{Hubble Type\tablenotemark{a}}
&\colhead{Heliocentric Radial Velocity\tablenotemark{a}}
&\colhead{Observed Supernovae}
&\colhead{$\Delta m_{15} (B)$}\\
\colhead{}
&\colhead{}
&\colhead{(km/s)}}
\startdata
NGC 524  & SA(rs)0+ & 2379 & 2000cx(Ia)    & 0.93 $\pm$ 0.04\tablenotemark{b}\\
NGC 1316 & SAB(s)& 1760 & 1980N(Ia), 1981D(Ia) & 
1.28 $\pm$ 0.04\tablenotemark{c}, 1.25 $\pm$ 0.15\tablenotemark{d}\\ 
NGC 1380 & SA0   & 1877 & 1992A(Ia)        & 1.47 $\pm$ 0.05\tablenotemark{c}\\
NGC 1448 & SAcd: sp 
& 1168 & 1983S(II), 2001e1(Ia), 2003hn(II) & --, 
1.13 $\pm$ 0.04\tablenotemark{e}, -- \\
NGC 4526 & SAB(s)00 & 612  \tablenotemark{f} 
& 1969E, 1994D(Ia) & --, 1.32 $\pm$ 0.05
\tablenotemark{c}\\
\enddata
\tablenotetext{a}{Data taken from the NASA/IPAC Extragalactic Database}
\tablenotetext{b}{Data taken from \citet{li2001}}
\tablenotetext{c}{Data taken from \citet{phillips1999}}
\tablenotetext{d}{Phillips, 2006, private communication}
\tablenotetext{e}{Data taken from \citet{kris2003}}
\tablenotetext{f}{Data taken from \citet{rubin1999}}
\end{deluxetable}

\begin{deluxetable}{lclccc}
\tablenum{2}
\tabletypesize{\footnotesize}
\tablewidth{0pt}
\tablecaption{Observing Log\label{table:log}}
\tablehead{
\colhead{Galaxy}
&\colhead{Dates Observed}
&\colhead{Telescope}
&\colhead{Total Onband}
&\colhead{Total Offband}
&\colhead{Mean Seeing}\\
\colhead{}
&\colhead{}
&\colhead{}
&\colhead{Exposure time (hours)}
&\colhead{Exposure time (hours)}
&\colhead{}
}
\startdata
NGC~524  & Dec. 18-19, 2003  & Clay 6.5m & 3.75 & 0.33 & $0\parcsec7$\\
NGC~1316 & Dec. 18-19, 2003  & Clay 6.5m & 3.75 & 0.50 & $0\parcsec8$\\
NGC~1380 & Dec. 18-19, 2003  & Clay 6.5m & 4.00 & 0.50 & $0\parcsec6$\\
NGC~1448 & Nov. 11-12, 2004  & Clay 6.5m & 2.50 & 0.33 & $0\parcsec8$\\
NGC~4526 & Mar. 10-11, 2005  & WIYN 3.5m & 6.00 & 1.50 & $0\parcsec8$\\
\enddata
\end{deluxetable}

\begin{deluxetable}{lccccc}
\tablenum{3}
\tablewidth{0pt}
\tablecaption{NGC 1316 Planetary Nebula Candidates\label{table:n1316pn}}
\tablehead{
\colhead{ID} & \colhead{Name} &\colhead{$\alpha(2000)$} &\colhead{$\delta(2000)$} 
&\colhead{$m_{5007}$} 
&\colhead{Notes\tablenotemark{a}}\\
}
\startdata
 1   & PNE N1316 J032238.08-371310.22    & 3 22 38.08 & -37 13 10.22  & 26.789 & S\\
 2   & PNE N1316 J032235.47-371350.81    & 3 22 35.47 & -37 13 50.81  & 26.840 & S\\
 3   & PNE N1316 J032233.50-371318.51    & 3 22 33.50 & -37 13 18.51  & 26.871 & S\\
 4   & PNE N1316 J032235.98-371405.18    & 3 22 35.98 & -37 14 05.18  & 26.901 & S\\
 5   & PNE N1316 J032234.97-371404.68    & 3 22 34.97 & -37 14 04.68  & 26.903 & S\\
 6   & PNE N1316 J032234.55-371352.46    & 3 22 34.55 & -37 13 52.46  & 26.904 & S\\
 7   & PNE N1316 J032232.06-371255.46    & 3 22 32.06 & -37 12 55.46  & 26.931 & S\\
 8   & PNE N1316 J032237.32-371358.06    & 3 22 37.32 & -37 13 58.06  & 26.968 & S\\
 9   & PNE N1316 J032233.74-371352.24    & 3 22 33.74 & -37 13 52.24  & 26.973 & S\\
 10  & PNE N1316 J032239.91-371438.52    & 3 22 39.91 & -37 14 38.52  & 26.993 & S\\
 11  & PNE N1316 J032230.69-371408.43    & 3 22 30.69 & -37 14 08.43  & 26.998 & S\\
 12  & PNE N1316 J032237.72-371338.00    & 3 22 37.72 & -37 13 38.00  & 27.007 & S\\
 13  & PNE N1316 J032236.91-371413.01    & 3 22 36.91 & -37 14 13.01  & 27.012 & S, MCJ78\\
 14  & PNE N1316 J032240.02-371425.61    & 3 22 40.02 & -37 14 25.61  & 27.028 & S, MCJ51\\
 15  & PNE N1316 J032236.36-371431.96    & 3 22 36.36 & -37 14 31.96  & 27.060 & S\\
 16  & PNE N1316 J032233.01-371405.43    & 3 22 33.01 & -37 14 05.43  & 27.083 & S, MCJ87\\
 17  & PNE N1316 J032236.30-371255.58    & 3 22 36.30 & -37 12 55.58  & 27.098 & S\\
 18  & PNE N1316 J032239.92-371443.34    & 3 22 39.92 & -37 14 43.34  & 27.101 & S\\
 19  & PNE N1316 J032240.65-371440.10    & 3 22 40.65 & -37 14 40.10  & 27.121 & S\\
 20  & PNE N1316 J032237.45-371414.68    & 3 22 37.45 & -37 14 14.68  & 27.135 & S\\
 21  & PNE N1316 J032234.10-371334.24    & 3 22 34.10 & -37 13 34.24  & 27.135 & S\\
 22  & PNE N1316 J032231.38-371413.82    & 3 22 31.38 & -37 14 13.82  & 27.143 & S\\
 23  & PNE N1316 J032234.02-371251.02    & 3 22 34.02 & -37 12 51.02  & 27.150 & S\\
 24  & PNE N1316 J032232.10-371315.74    & 3 22 32.10 & -37 13 15.74  & 27.158 & S\\
 25  & PNE N1316 J032238.26-371438.19    & 3 22 38.26 & -37 14 38.19  & 27.161 & S\\
 26  & PNE N1316 J032233.38-371310.89    & 3 22 33.38 & -37 13 10.89  & 27.184 & S\\
 27  & PNE N1316 J032230.39-371433.06    & 3 22 30.39 & -37 14 33.06  & 27.191 & S, MCJ18\\
 28  & PNE N1316 J032236.72-371323.25    & 3 22 36.72 & -37 13 23.25  & 27.204 & \\
 29  & PNE N1316 J032237.32-371356.43    & 3 22 37.32 & -37 13 56.43  & 27.213 & \\
 30  & PNE N1316 J032235.24-371420.97    & 3 22 35.24 & -37 14 20.97  & 27.213 & \\
 31  & PNE N1316 J032236.33-371413.05    & 3 22 36.33 & -37 14 13.05  & 27.264 & \\
 32  & PNE N1316 J032236.60-371409.57    & 3 22 36.60 & -37 14 09.57  & 27.279 & \\
 33  & PNE N1316 J032232.87-371302.21    & 3 22 32.87 & -37 13 02.21  & 27.302 & \\
 34  & PNE N1316 J032236.48-371439.48    & 3 22 36.48 & -37 14 39.48  & 27.305 & \\
 35  & PNE N1316 J032230.94-371347.74    & 3 22 30.94 & -37 13 47.74  & 27.339 & \\
 36  & PNE N1316 J032231.21-371345.42    & 3 22 31.21 & -37 13 45.42  & 27.341 & \\
 37  & PNE N1316 J032235.16-371353.53    & 3 22 35.16 & -37 13 53.53  & 27.343 & \\
 38  & PNE N1316 J032233.31-371333.17    & 3 22 33.31 & -37 13 33.17  & 27.349 & \\
 39  & PNE N1316 J032240.25-371411.77    & 3 22 40.25 & -37 14 11.77  & 27.354 & \\
 40  & PNE N1316 J032237.07-371322.10    & 3 22 37.07 & -37 13 22.10  & 27.365 & \\
 41  & PNE N1316 J032233.14-371252.12    & 3 22 33.14 & -37 12 52.12  & 27.437 & \\
 42  & PNE N1316 J032239.12-371413.46    & 3 22 39.12 & -37 14 13.46  & 27.478 & \\
 43  & PNE N1316 J032238.69-371409.93    & 3 22 38.69 & -37 14 09.93  & 27.514 & \\
 44  & PNE N1316 J032239.43-371352.83    & 3 22 39.43 & -37 13 52.83  & 27.524 & \\
 45  & PNE N1316 J032232.79-371259.84    & 3 22 32.79 & -37 12 59.84  & 27.584 & \\
\enddata
\tablenotetext{a}{The letter ``S'' indicates that the PN candidate is part
of the photometrically complete sub-sample.  The ``MCJ'' designation is
a PN candidate originally detected by \citet{mcmillan1993}, with the number
referring to the original identification number.}
\end{deluxetable}

\clearpage

\begin{deluxetable}{lccccc}
\tablenum{4}
\tablewidth{0pt}
\tablecaption{NGC 1380 Planetary Nebula Candidates\label{table:n1380pn}}
\tablehead{
\colhead{ID} & \colhead{Name} &\colhead{$\alpha(2000)$} &\colhead{$\delta(2000)$} 
&\colhead{$m_{5007}$} 
&\colhead{Notes\tablenotemark{a}}\\
}
\startdata
1     & PNE N1380 J033627.64-345759.86 &  3 36 27.64&  -34 57 59.86 &  26.585 & S\\   
2     & PNE N1380 J033628.13-345754.00 &  3 36 28.13&  -34 57 54.00 &  26.718 & S\\   
3     & PNE N1380 J033628.09-345815.27 &  3 36 28.09&  -34 58 15.27 &  26.725 & S\\   
4     & PNE N1380 J033629.32-345719.12 &  3 36 29.32&  -34 57 19.12 &  26.770 & S\\   
5     & PNE N1380 J033626.87-345701.01 &  3 36 26.87&  -34 57 01.01 &  26.776 & S\\   
6     & PNE N1380 J033631.78-345805.89 &  3 36 31.78&  -34 58 05.89 &  26.785 & S\\   
7     & PNE N1380 J033626.48-345830.20 &  3 36 26.48&  -34 58 30.20 &  26.809 & S\\   
8     & PNE N1380 J033629.89-345758.44 &  3 36 29.89&  -34 57 58.44 &  26.821 & S\\   
9     & PNE N1380 J033628.40-345742.52 &  3 36 28.40&  -34 57 42.52 &  26.952 & S\\   
10    & PNE N1380 J033624.02-345618.03 &  3 36 24.02&  -34 56 18.03 &  26.977 & S\\   
11    & PNE N1380 J033631.75-345833.44 &  3 36 31.75&  -34 58 33.44 &  27.011 & S\\   
12    & PNE N1380 J033631.34-345800.91 &  3 36 31.34&  -34 58 00.91 &  27.040 & S\\   
13    & PNE N1380 J033629.15-345806.29 &  3 36 29.15&  -34 58 06.29 &  27.079 & S\\   
14    & PNE N1380 J033625.86-345729.68 &  3 36 25.86&  -34 57 29.68 &  27.079 & S\\   
15    & PNE N1380 J033625.17-345800.55 &  3 36 25.17&  -34 58 00.55 &  27.079 & S\\   
16    & PNE N1380 J033632.84-345759.34 &  3 36 32.84&  -34 57 59.34 &  27.097 & S\\   
17    & PNE N1380 J033629.24-345739.72 &  3 36 29.24&  -34 57 39.72 &  27.107 & S\\   
18    & PNE N1380 J033623.04-345656.01 &  3 36 23.04&  -34 56 56.01 &  27.163 & S\\   
19    & PNE N1380 J033626.71-345754.48 &  3 36 26.71&  -34 57 54.48 &  27.188 & S\\   
20    & PNE N1380 J033627.72-345740.67 &  3 36 27.72&  -34 57 40.67 &  27.201 & \\   
21    & PNE N1380 J033625.72-345819.32 &  3 36 25.72&  -34 58 19.32 &  27.203 & \\   
22    & PNE N1380 J033624.90-345809.44 &  3 36 24.90&  -34 58 09.44 &  27.205 & \\   
23    & PNE N1380 J033624.01-345740.16 &  3 36 24.01&  -34 57 40.16 &  27.209 & \\   
24    & PNE N1380 J033622.25-345832.15 &  3 36 22.25&  -34 58 32.15 &  27.213 & \\   
25    & PNE N1380 J033632.25-345810.38 &  3 36 32.25&  -34 58 10.38 &  27.289 & \\   
26    & PNE N1380 J033627.31-345727.95 &  3 36 27.31&  -34 57 27.95 &  27.321 & \\   
27    & PNE N1380 J033626.65-345618.90 &  3 36 26.65&  -34 56 18.90 &  27.361 & \\   
28    & PNE N1380 J033624.92-345748.34 &  3 36 24.92&  -34 57 48.34 &  27.386 & \\   
29    & PNE N1380 J033628.19-345718.43 &  3 36 28.19&  -34 57 18.43 &  27.399 & \\   
30    & PNE N1380 J033625.88-345728.05 &  3 36 25.88&  -34 57 28.05 &  27.411 & \\   
31    & PNE N1380 J033631.69-345648.84 &  3 36 31.69&  -34 56 48.84 &  27.413 & \\   
32    & PNE N1380 J033625.59-345706.14 &  3 36 25.59&  -34 57 06.14 &  27.435 & \\   
33    & PNE N1380 J033626.50-345729.23 &  3 36 26.50&  -34 57 29.23 &  27.489 & \\   
34    & PNE N1380 J033627.11-345728.06 &  3 36 27.11&  -34 57 28.06 &  27.528 & \\   
35    & PNE N1380 J033624.14-345740.54 &  3 36 24.14&  -34 57 40.54 &  27.654 & \\   
36    & PNE N1380 J033623.46-345808.85 &  3 36 23.46&  -34 58 08.85 &  27.731 & \\   
37    & PNE N1380 J033630.00-345720.97 &  3 36 30.00&  -34 57 20.97 &  27.747 & \\   
38    & PNE N1380 J033630.06-345735.60 &  3 36 30.06&  -34 57 35.60 &  27.760 & \\   
39    & PNE N1380 J033624.14-345752.89 &  3 36 24.14&  -34 57 52.89 &  27.774 & \\   
40    & PNE N1380 J033624.64-345726.64 &  3 36 24.64&  -34 57 26.64 &  27.810 & \\   
41    & PNE N1380 J033629.73-345725.35 &  3 36 29.73&  -34 57 25.35 &  27.812 & \\   
42    & PNE N1380 J033622.56-345651.60 &  3 36 22.56&  -34 56 51.60 &  27.839 & \\   
43    & PNE N1380 J033624.99-345623.17 &  3 36 24.99&  -34 56 23.17 &  27.941 & \\   
44    & PNE N1380 J033626.14-345646.42 &  3 36 26.14&  -34 56 46.42 &  28.223 & \\   
\enddata
\tablenotetext{a}{The letter ``S'' indicates that the PN candidate is part
of the photometrically complete sub-sample.}
\end{deluxetable}

\LongTables
\begin{deluxetable}{lccccc}
\tablenum{5}
\tablewidth{0pt}
\tablecaption{NGC 4526 Planetary Nebula Candidates\label{table:n4526pn}}
\tablehead{
\colhead{ID} & \colhead{Name} &\colhead{$\alpha(2000)$} &\colhead{$\delta(2000)$} 
&\colhead{$m_{5007}$} 
&\colhead{Notes}\\
}
\startdata
 1     & PNE N4526 J123351.98074112.94  & 12 33 51.98   & 7 41 12.94  & 26.101 & S,O \\
 2     & PNE N4526 J123353.40074230.79  & 12 33 53.40   & 7 42 30.79  & 26.271 & S,O \\
 3     & PNE N4526 J123402.50074143.67  & 12 34 02.50   & 7 41 43.67  & 26.274 & S,I \\
 4     & PNE N4526 J123353.85074238.88  & 12 33 53.85   & 7 42 38.88  & 26.316 & S,O \\
 5     & PNE N4526 J123404.61074215.67  & 12 34 04.61   & 7 42 15.67  & 26.375 & S,I \\
 6     & PNE N4526 J123400.54074234.82  & 12 34 00.54   & 7 42 34.82  & 26.402 & S,I \\
 7     & PNE N4526 J123354.25074236.83  & 12 33 54.25   & 7 42 36.83  & 26.435 & S,O \\
 8     & PNE N4526 J123359.74074157.58  & 12 33 59.74   & 7 41 57.58  & 26.440 & S,I \\
 9     & PNE N4526 J123405.27074155.36  & 12 34 05.27   & 7 41 55.36  & 26.464 & S,I \\
 10    & PNE N4526 J123357.14074328.80  & 12 33 57.14   & 7 43 28.80  & 26.472 & S,O \\
 11    & PNE N4526 J123348.82074045.49  & 12 33 48.82   & 7 40 45.49  & 26.476 & S,O \\
 12    & PNE N4526 J123358.16074207.65  & 12 33 58.16   & 7 42 07.65  & 26.509 & S,I \\
 13    & PNE N4526 J123403.45074130.96  & 12 34 03.45   & 7 41 30.96  & 26.510 & S,I \\
 14    & PNE N4526 J123354.73074258.81  & 12 33 54.73   & 7 42 58.81  & 26.514 & S,O \\
 15    & PNE N4526 J123401.68074217.53  & 12 34 01.68   & 7 42 17.53  & 26.516 & S,I \\
 16    & PNE N4526 J123356.47074130.23  & 12 33 56.47   & 7 41 30.23  & 26.548 & S,O \\
 17    & PNE N4526 J123402.07074226.38  & 12 34 02.07   & 7 42 26.38  & 26.559 & S,I \\
 18    & PNE N4526 J123352.67074152.67  & 12 33 52.67   & 7 41 52.67  & 26.582 & S,O \\
 19    & PNE N4526 J123355.54074235.11  & 12 33 55.54   & 7 42 35.11  & 26.584 & S,O \\
 20    & PNE N4526 J123403.30074133.96  & 12 34 03.30   & 7 41 33.96  & 26.589 & S,I \\
 21    & PNE N4526 J123352.76074228.76  & 12 33 52.76   & 7 42 28.76  & 26.590 & S,O \\
 22    & PNE N4526 J123357.55074236.42  & 12 33 57.55   & 7 42 36.42  & 26.599 & S,I \\
 23    & PNE N4526 J123403.36074140.87  & 12 34 03.36   & 7 41 40.87  & 26.618 & S,I \\
 24    & PNE N4526 J123358.03074040.17  & 12 33 58.03   & 7 40 40.17  & 26.642 & S,O \\
 25    & PNE N4526 J123406.16074140.70  & 12 34 06.16   & 7 41 40.70  & 26.700 & S,I \\
 26    & PNE N4526 J123401.76074250.59  & 12 34 01.76   & 7 42 50.59  & 26.732 & S,I \\
 27    & PNE N4526 J123356.19074231.95  & 12 33 56.19   & 7 42 31.95  & 26.734 & S,I \\
 28    & PNE N4526 J123402.31074247.35  & 12 34 02.31   & 7 42 47.35  & 26.758 & S,I \\
 29    & PNE N4526 J123358.51074159.04  & 12 33 58.51   & 7 41 59.04  & 26.776 & S,I \\
 30    & PNE N4526 J123402.51074123.22  & 12 34 02.51   & 7 41 23.22  & 26.798 & S,I \\
 31    & PNE N4526 J123402.96074219.46  & 12 34 02.96   & 7 42 19.46  & 26.810 & S,I \\
 32    & PNE N4526 J123401.47074156.06  & 12 34 01.47   & 7 41 56.06  & 26.838 & S,I \\
 33    & PNE N4526 J123406.57074119.76  & 12 34 06.57   & 7 41 19.76  & 26.839 & S,I \\
 34    & PNE N4526 J123359.95074149.63  & 12 33 59.95   & 7 41 49.63  & 26.843 & S,I \\
 35    & PNE N4526 J123401.24074140.93  & 12 34 01.24   & 7 41 40.93  & 26.855 & S,I \\
 36    & PNE N4526 J123357.14074258.81  & 12 33 57.14   & 7 42 58.81  & 26.871 & S,I \\
 37    & PNE N4526 J123357.81074237.94  & 12 33 57.81   & 7 42 37.94  & 26.878 & S,I \\
 38    & PNE N4526 J123355.17074203.61  & 12 33 55.17   & 7 42 03.61  & 26.888 & S,O \\
 39    & PNE N4526 J123401.66074139.78  & 12 34 01.66   & 7 41 39.78  & 26.891 & S,I \\
 40    & PNE N4526 J123402.78074136.58  & 12 34 02.78   & 7 41 36.58  & 26.896 & S,I \\
 41    & PNE N4526 J123400.05074302.28  & 12 34 00.05   & 7 43 02.28  & 26.905 & S,I \\
 42    & PNE N4526 J123350.77074332.74  & 12 33 50.77   & 7 43 32.74  & 26.915 & S,O \\
 43    & PNE N4526 J123400.64074209.43  & 12 34 00.64   & 7 42 09.43  & 26.919 & S,I \\
 44    & PNE N4526 J123400.10074221.54  & 12 34 00.10   & 7 42 21.54  & 26.924 & S,I \\
 45    & PNE N4526 J123401.99074250.13  & 12 34 01.99   & 7 42 50.13  & 26.939 & S,I \\
 46    & PNE N4526 J123406.75074138.19  & 12 34 06.75   & 7 41 38.19  & 26.941 & S,I \\
 47    & PNE N4526 J123403.35074138.00  & 12 34 03.35   & 7 41 38.00  & 26.941 & S,I \\
 48    & PNE N4526 J123351.81074301.43  & 12 33 51.81   & 7 43 01.43  & 26.958 & S,O \\
 49    & PNE N4526 J123404.95074211.16  & 12 34 04.95   & 7 42 11.16  & 26.964 & S,I \\
 50    & PNE N4526 J123404.39074104.24  & 12 34 04.39   & 7 41 04.24  & 26.965 & S,I \\
 51    & PNE N4526 J123405.48074120.59  & 12 34 05.48   & 7 41 20.59  & 27.000 & S,I \\
 52    & PNE N4526 J123405.22074158.22  & 12 34 05.22   & 7 41 58.22  & 27.006 & S,I \\
 53    & PNE N4526 J123354.57074219.45  & 12 33 54.57   & 7 42 19.45  & 27.043 & S,O \\
 54    & PNE N4526 J123408.99074152.18  & 12 34 08.99   & 7 41 52.18  & 27.053 & S,I \\
 55    & PNE N4526 J123351.37074312.81  & 12 33 51.37   & 7 43 12.81  & 27.054 & S,O \\
 56    & PNE N4526 J123402.60074223.62  & 12 34 02.60   & 7 42 23.62  & 27.071 & S,I \\
 57    & PNE N4526 J123351.09074234.56  & 12 33 51.09   & 7 42 34.56  & 27.093 & S,O \\
 58    & PNE N4526 J123355.98074250.73  & 12 33 55.98   & 7 42 50.73  & 27.094 & S,O \\
 59    & PNE N4526 J123350.54074219.91  & 12 33 50.54   & 7 42 19.91  & 27.094 & S,O \\
 60    & PNE N4526 J123359.34074229.09  & 12 33 59.34   & 7 42 29.09  & 27.120 & I \\
 61    & PNE N4526 J123359.29074111.18  & 12 33 59.29   & 7 41 11.18  & 27.126 & O \\
 62    & PNE N4526 J123356.34074142.48  & 12 33 56.34   & 7 41 42.48  & 27.136 & O \\
 63    & PNE N4526 J123400.42074152.78  & 12 34 00.42   & 7 41 52.78  & 27.147 & I \\
 64    & PNE N4526 J123355.57074028.60  & 12 33 55.57   & 7 40 28.60  & 27.161 & O \\
 65    & PNE N4526 J123351.44074302.47  & 12 33 51.44   & 7 43 02.47  & 27.183 & O \\
 66    & PNE N4526 J123400.69074250.15  & 12 34 00.69   & 7 42 50.15  & 27.206 & I \\
 67    & PNE N4526 J123356.28074143.29  & 12 33 56.28   & 7 41 43.29  & 27.217 & O \\
 68    & PNE N4526 J123403.70074233.44  & 12 34 03.70   & 7 42 33.44  & 27.241 & I \\
 69    & PNE N4526 J123358.64074216.55  & 12 33 58.64   & 7 42 16.55  & 27.246 & I \\
 70    & PNE N4526 J123351.09074103.07  & 12 33 51.09   & 7 41 03.07  & 27.268 & O \\
 71    & PNE N4526 J123349.95074238.34  & 12 33 49.95   & 7 42 38.34  & 27.273 & O \\
 72    & PNE N4526 J123352.27074325.66  & 12 33 52.27   & 7 43 25.66  & 27.274 & O \\
 73    & PNE N4526 J123350.73074112.66  & 12 33 50.73   & 7 41 12.66  & 27.283 & O \\
 74    & PNE N4526 J123358.85074214.49  & 12 33 58.85   & 7 42 14.49  & 27.289 & I \\
 75    & PNE N4526 J123355.26074228.69  & 12 33 55.26   & 7 42 28.69  & 27.293 & O \\
 76    & PNE N4526 J123356.76074228.45  & 12 33 56.76   & 7 42 28.45  & 27.417 & I \\
 77    & PNE N4526 J123354.07074320.01  & 12 33 54.07   & 7 43 20.01  & 27.427 & O \\
 78    & PNE N4526 J123402.45074233.94  & 12 34 02.45   & 7 42 33.94  & 27.434 & I \\
 79    & PNE N4526 J123356.99074053.37  & 12 33 56.99   & 7 40 53.37  & 27.442 & O \\
 80    & PNE N4526 J123358.01074132.05  & 12 33 58.01   & 7 41 32.05  & 27.446 & O \\
 81    & PNE N4526 J123352.43074209.19  & 12 33 52.43   & 7 42 09.19  & 27.462 & O \\
 82    & PNE N4526 J123405.22074318.03  & 12 34 05.22   & 7 43 18.03  & 27.471 & O \\
 83    & PNE N4526 J123409.09074221.56  & 12 34 09.09   & 7 42 21.56  & 27.500 & O \\
 84    & PNE N4526 J123404.55074046.17  & 12 34 04.55   & 7 40 46.17  & 27.501 & O \\
 85    & PNE N4526 J123405.32074036.85  & 12 34 05.32   & 7 40 36.85  & 27.551 & O \\
 86    & PNE N4526 J123348.89074158.37  & 12 33 48.89   & 7 41 58.37  & 27.600 & O \\
 87    & PNE N4526 J123359.69074308.94  & 12 33 59.69   & 7 43 08.94  & 27.620 & O \\
 88    & PNE N4526 J123356.18074059.98  & 12 33 56.18   & 7 40 59.98  & 27.648 & O \\
 89    & PNE N4526 J123404.15074027.15  & 12 34 04.15   & 7 40 27.15  & 27.652 & O \\
 90    & PNE N4526 J123402.42074305.87  & 12 34 02.42   & 7 43 05.87  & 27.678 & O \\
 91    & PNE N4526 J123406.88074025.57  & 12 34 06.88   & 7 40 25.57  & 27.696 & O \\
 92    & PNE N4526 J123358.26074022.60  & 12 33 58.26   & 7 40 22.60  & 27.763 & O \\
 93    & PNE N4526 J123353.00074204.12  & 12 33 53.00   & 7 42 04.12  & 27.791 & O \\
 94    & PNE N4526 J123345.66074309.87  & 12 33 45.66   & 7 43 09.87  & 27.990 & O \\
\enddata
\tablenotetext{a}{The letter ``S'' indicates that the PN candidate is part
of the photometrically complete sub-sample.  The letter ``I'' or
``O'' indicates whether the object is part of the inner or outer sub-sample
discussed in the text.}
\end{deluxetable}

\begin{deluxetable}{lcccccc}
\tablenum{6}
\tablewidth{0pt}
\tablecaption{PN Photometric Error versus Magnitude\label{table:error}}
\tablehead{
\colhead{m$_{5007}$}
&\colhead{NGC~1316}
&\colhead{NGC~1316}
&\colhead{NGC~1380}
&\colhead{NGC~1380}
&\colhead{NGC~4526}
&\colhead{NGC~4526}\\
\colhead{}
&\colhead{Mean 1$\sigma$ error}
&\colhead{Number}
&\colhead{Mean 1$\sigma$ error}
&\colhead{Number}
&\colhead{Mean 1$\sigma$ error}
&\colhead{Number}
}
\startdata
26.3 &       &    &       &    & 0.063 &  3\\
26.5 &       &    &       &    & 0.070 &  8\\
26.7 & 0.087 &  1 & 0.094 &  5 & 0.085 & 14\\ 
26.9 & 0.086 & 10 & 0.109 &  4 & 0.101 & 12\\
27.1 & 0.098 & 16 & 0.107 &  9 & 0.117 & 19\\
27.3 & 0.130 & 13 & 0.127 & 10 & 0.120 & 17\\
27.5 & 0.140 &  5 & 0.139 &  5 & 0.138 &  9\\
27.7 &       &    & 0.176 &  5 & 0.159 &  8\\
27.9 &       &    & 0.191 &  4 & 0.175 &  3\\
28.1 &       &    &       &    & 0.188 &  1\\
28.3 &       &    & 0.265 &  1 & 0.441 &  1\\
\enddata
\end{deluxetable}

\begin{deluxetable}{lccl}
\tablenum{7}
\tablewidth{0pt}
\tablecaption{Comparison of NGC~1316 PN magnitudes\label{table:compare}}
\tablehead{
\colhead{ID}
&\colhead{m$_{5007}$ (MCJ1993)}
&\colhead{m$_{5007}$ (this work)}
&\colhead{$\Delta m_{5007}$}
}
\startdata
MCJ12  / --   & 26.858 $\pm$ 0.10 & 26.940 $\pm$ 0.08 & -0.081 $\pm$ 0.129\\ 
MCJ16  / --   & 26.927 $\pm$ 0.12 & 27.059 $\pm$ 0.10 & -0.132 $\pm$ 0.154\\
MCJ18  / 27   & 26.943 $\pm$ 0.12 & 27.191 $\pm$ 0.11 & -0.248 $\pm$ 0.163\\
MCJ48  / --   & 27.086 $\pm$ 0.13 & 27.118 $\pm$ 0.13 & -0.032 $\pm$ 0.184\\
MCJ51  / 14   & 27.141 $\pm$ 0.13 & 27.028 $\pm$ 0.09 &  0.113 $\pm$ 0.157\\
MCJ63  / --   & 27.204 $\pm$ 0.14 & 27.318 $\pm$ 0.12 & -0.114 $\pm$ 0.182\\
MCJ78  / 13   & 27.277 $\pm$ 0.14 & 27.012 $\pm$ 0.09 &  0.265 $\pm$ 0.166\\
MCJ87  / 16   & 27.351 $\pm$ 0.15 & 27.083 $\pm$ 0.10 &  0.268 $\pm$ 0.179\\
Weighted Mean & & & -0.0053 $\pm$ 0.057\\
\enddata
\end{deluxetable}

\begin{deluxetable}{llllllllll}
\tablenum{8}
\tablewidth{0pt}
\tabletypesize{\footnotesize}
\tablecaption{SNe Ia with Cepheid, SBF, or PNLF Distances\label{table:distances}}
\tablehead{
\colhead{Host}
&\colhead{SN}
&\colhead{$\Delta~m_{15}(B)$}
&\colhead{$(m-M)_0$}
&\colhead{ref.}
&\colhead{$(m-M)_0$}
&\colhead{ref.}
&\colhead{$(m-M)_0$}
&\colhead{ref.}
&\colhead{Calib.?}\\
\colhead{}
& \colhead{}
& \colhead{}
& \colhead{Cepheid}
& \colhead{}
& \colhead{SBF}
& \colhead{}
& \colhead{PNLF}
& \colhead{}
&\colhead{}
}
\startdata
NGC 5253    & 1972E   & 0.88 (10)  & 27.56 (14)  & a       & -           & -   & $28.03^{+.08}_{-.65}$  & f &      -\\
NGC 1316    & 1980N   & 1.28 (04)  &     -       & -       & 31.50 (14)  & e   & $31.26^{+.09}_{-.12}$  & g &   opt+NIR\\
NGC 1316    & 1981D   & 1.25 (15)  &     -       & -       & 31.50 (14)  & e   & $31.26^{+.09}_{-.12}$  & g &    \\
NGC 4536    & 1981B   & 1.11 (07)  & 30.80 (04)  & a       & -           & -   &        -               & - &   opt+NIR\\
NGC 5128    & 1986G   & 1.81 (07)  &     -       & -       & 27.96 (14)  & e   & $27.64^{+.09}_{-.09}$  & f &      -\\
NGC 3627    & 1989B   & 1.35 (07)  & 29.86 (08)  & a       & -           & -   & $29.90^{+.07}_{-.09}$  & f &    \\
NGC 4639    & 1990N   & 1.08 (05)  & 31.61 (08)  & a       & -           & -   &        -               & - &     opt\\
NGC 4527    & 1991T   & 0.96 (05)  & 30.53 (09)  & b       & -           & -   &        -               & - &      -\\
NGC 4374    & 1991bg  & 1.94 (10)  &     -       & -       & 31.16 (11)  & e   & $30.89^{+.09}_{-.11}$  & f &      -\\
NGC 1380    & 1992A   & 1.47 (05)  &     -       & -       & 31.07 (18)  & e   & $31.04^{+.11}_{-.15}$  & g &     opt\\
ESO 352-G57 & 1992bo  & 1.69 (05)  &     -       & -       & 34.17 (15)  & e   &        -               & - &     opt\\
NGC 4526    & 1994D   & 1.32 (05)  &     -       & -       & 30.98 (20)  & e   & $30.66^{+.20}_{-.20}$  & g &   opt+NIR\\
NGC 3370    & 1994ae  & 1.02 (10)  & 32.13 (03)  & c       & -           & -   &        -               & - &     opt\\
NGC 2962    & 1995D   & 1.00 (05)  &     -       & -       & 32.50 (15)  & e   &        -               & - &     opt\\
NGC 5061    & 1996X   & 1.26 (05)  &     -       & -       & 32.16 (19)  & e   &        -               & - &     opt\\
NGC 5308    & 1996bk  & 1.78 (10)  &     -       & -       & 32.39 (21)  & e   &        -               & - &      -\\
NGC 3982    & 1998aq  & 1.16 (10)  & 31.56 (08)  & c       & -           & -   &        -               & - &     opt\\
NGC 6495    & 1998bp  & 1.96 (10)  &     -       & -       & 33.00 (15)  & e   &        -               & - &      -\\
NGC 3368    & 1998bu  & 1.05 (05)  & 29.97 (06)  & a       & -           & -   & $29.79^{+.08}_{-.10}$  & f &     NIR\\
NGC 2841    & 1999by  & 1.90 (05)  & 30.58 (06)  & d       & -           & -   &        -               & - &      -\\
\enddata
\tablerefs{(a) Freedman et al. (2001); (b) Gibson \& Stetson (2001); (c) - Riess et al. (2005); (d) Macri et al. (2001);
(e) - Tonry et al. (2001), increased by 0.12 mag as per Jensen et al. (2003); (f) - Ciardullo et al. (2002), Tables 9 \& 10 
(g) - This paper}
\end{deluxetable}

\begin{deluxetable}{lcccccccc}
\tablenum{9}
\tablewidth{0pt}
\tabletypesize{\footnotesize}
\tablecaption{Absolute magnitudes of SNe~Ia derived from the PNLF\label{table:absmag}}
\tablehead{
\colhead{SN}
&\colhead{Host}
&\colhead{$\Delta~m_{15}(B)$}
&\colhead{M$_{0}$(B)}
&\colhead{M$_{0}$(V)}
&\colhead{M$_{0}$(I)}
&\colhead{M$_{0}$(J)}
&\colhead{M$_{0}$(H)}
&\colhead{M$_{0}$(K)}
}
\startdata
1980N  & NGC 1316  & 1.29(04)  & -19.09(16) & -19.06(15) & -18.70(14) & -18.48(14) & -18.06(16) & -18.18(16)\\
1992A  & NGC 1380  & 1.47(05)  & -18.52(19) & -18.50(17) & -18.21(16) & ...        &  ...       & ...\\
1994D  & NGC 4526  & 1.32(05)  & -18.83(23) & -18.78(22) & -18.52(21) & -18.04(22) & -18.03(22) &-18.08(22)\\
1998bu & NGC 3368  & 1.05(05)  & --         & --         & --         & -18.46(12) & -18.23(14) &-18.37(10)\\
\enddata
\end{deluxetable}

\begin{deluxetable}{lccccc}
\tablenum{10}
\tablewidth{0pt}
\tabletypesize{\footnotesize}
\tablecaption{Hubble Constants from BVI Light Curves of SNe~Ia\label{table:hubblebvi}}
\tablehead{
\colhead{Method}
&\colhead{Number of Calibrators}
&\colhead{B\tablenotemark{a}}
&\colhead{V\tablenotemark{a}}
&\colhead{I\tablenotemark{a}}
&\colhead{Average\tablenotemark{a}}\\
\colhead{} 
&\colhead{}
&\colhead{(\ho)} 
&\colhead{(\ho)} 
&\colhead{(\ho)} 
&\colhead{(\ho)}
}
\startdata
Cepheid    & 4      & 75.5 (3.6)  & 75.9 (3.5)  & 75.3 (3.5)   & 75.6 (3.2)\\
SBF        & 6      & 76.5 (3.2)  & 76.7 (3.1)  & 76.1 (3.1)   & 76.4 (2.8)\\
PNLF       & 3      & 84.1 (4.5)  & 83.7 (4.4)  & 83.8 (4.4)   & 83.8 (4.2)\\
\enddata
\tablenotetext{a}{The error bars convey internal errors only -- see the text for a detailed
explanation.}
\end{deluxetable}

\begin{deluxetable}{lccccc}
\tablenum{11}
\tablewidth{0pt}
\tabletypesize{\footnotesize}
\tablecaption{Hubble Constant from JHK Light Curves of SNe~Ia\label{table:hubblejhk}}
\tablehead{
\colhead{Method}
&\colhead{Number of Calibrators}
&\colhead{J\tablenotemark{a}}
&\colhead{H\tablenotemark{a}}
&\colhead{K\tablenotemark{a}}
&\colhead{Average\tablenotemark{a}}\\
\colhead{} & \colhead{}
&\colhead{(\ho)} &\colhead{(\ho)} &\colhead{(\ho)} &\colhead{(\ho)}
}
\startdata
Cepheid    & 2 & 77.3 (6.0) & 76.7 (6.1) & 71.5 (5.9) & 75.0 (5.4)\\
SBF        & 3 & 78.5 (6.4) & 74.7 (6.1) & 73.1 (6.4) & 75.4 (4.9)\\
PNLF       & 3 & 82.3 (6.1) & 79.1 (5.9) & 76.4 (6.0) & 79.2 (5.0)\\
\enddata
\tablenotetext{a}{The error bars convey internal errors only -- see the text for a detailed
explanation.}
\end{deluxetable}

\pagebreak

\begin{figure}
\figurenum{2}
\label{fig:images}
%\plotone{pnlf_2.eps}
\caption{Our final [O~III] images for the four galaxies observed 
at Magellan in this survey.  From top to bottom and left to right 
the galaxies are NGC~524, NGC~1316, NGC~1380, and NGC~1448.  The 
images are approximately $2\parcmin3$ square.  North is to the 
right, and East is at the bottom of each image.  The planetary 
nebulae candidates detected in NGC~1316 and NGC~1380 are 
shown as points.}  
\end{figure}

\begin{figure}
\figurenum{3}
\label{fig:n4526image}
%\plotone{pnlf_3.eps}
\caption{Our final [O~III] image for NGC~4526, taken with the WIYN telescope.  
The image is approximately  $5\parcmin6$ East-West by $4\parcmin8$ 
North-South.  
North is up and east is to the left of this image.  
The planetary nebulae candidates detected in NGC~4526
are shown as points, with the candidates from the inner sample
given as the darker points, while the candidates from the outer sample 
have a lighter greyscale.  See the text for further explanation.}  
\end{figure}

%\clearpage

\pagebreak

\pagebreak

\end{document}